\documentclass[british,english,final,journal,10pt,twocolumn]{IEEEtran}
\usepackage[T1]{fontenc}
\usepackage[latin9]{inputenc}
\usepackage{amsmath}
\usepackage{amssymb}
\usepackage{graphicx}

\makeatletter

\providecommand{\tabularnewline}{\\}
\newcommand{\lyxdot}{.}

\usepackage{url}
\makeatother

\usepackage{babel}

\begin{document}

\title{Model-Based Speech Enhancement in the Modulation Domain}

\author{Yu Wang, \textit{Member}, \textit{IEEE} and Mike Brookes, \textit{Member},
\textit{IEEE}\thanks{Yu Wang is with the Department of Engineering, University of Cambridge, Cambridge CB2 1PZ, U.K. (email: yw396@cam.ac.uk)\par 
 Mike Brookes is with the Department of Electrical and Electronic Engineering, Imperial College, London SW7 2AZ, U.K. (email: mike.brookes@imperial.ac.uk)}}
 
 \IEEEpubid{10.1109/TASLP.2017.2786863~\copyright~2018 IEEE.}
 
 \markboth{IEEE/ACM Transactions on Audio, Speech and Language Processing, Vol. 26, No. 3, March 2018}{}
 
\maketitle

\begin{abstract}
This paper presents an algorithm for modulation-domain speech enhancement
using a Kalman filter. The proposed estimator jointly models the estimated
dynamics of the spectral amplitudes of speech and noise to obtain
an MMSE estimation of the speech amplitude spectrum with the assumption
that the speech and noise are additive in the complex domain. In order
to include the dynamics of noise amplitudes with those of speech amplitudes,
we propose a statistical ``Gaussring'' model that comprises a mixture
of Gaussians whose centres lie in a circle on the complex plane. The
performance of the proposed algorithm is evaluated using the perceptual
evaluation of speech quality (PESQ) measure, segmental SNR (segSNR)
measure and short-time objective intelligibility (STOI) measure. For
speech quality measures, the proposed algorithm is shown to give a
consistent improvement over a wide range of SNRs when compared to
competitive algorithms. Speech recognition experiments also show that
the Gaussring model based algorithm performs well for two types of
noise. 
\end{abstract}


\global\long\def\mixweight{\epsilon}
\global\long\def\mixmean{o}
\global\long\def\mixvar{\Delta}

\global\long\def\gringjs{\tilde{g}}
\global\long\def\gringjn{\breve{g}}

\global\long\def\meanjj{\boldsymbol{\mu}^{\left(\gringjs,\,\gringjn\right)}}
\global\long\def\varjj{\boldsymbol{\Sigma}^{\left(\gringjs,\,\gringjn\right)}}
\global\long\def\meansqjj{\boldsymbol{\mu}_{\mathsf{sq}}^{\left(\gringjs,\,\gringjn\right)}}
\global\long\def\varsqjj{\boldsymbol{\Sigma}_{\mathsf{sq}}^{\left(\gringjs,\,\gringjn\right)}}

\global\long\def\sigmasqjjs{\tilde{\sigma}_{\mathsf{sq}}^{\left(\gringjs,\,\gringjn\right)}}
\global\long\def\sigmasqjjn{\breve{\sigma}_{\mathsf{sq}}^{\left(\gringjs,\,\gringjn\right)}}
\global\long\def\rhojj{\rho_{\mathsf{sq}}^{\left(\gringjs,\,\gringjn\right)}}

\section{Introduction}

\subsection{Statistical Models for Speech Enhancement}

A popular class of speech enhancement algorithm derives an optimal
estimator for the spectral amplitudes based on assumed statistical
models for the speech and noise amplitudes in the short-time Fourier
transform (STFT) domain \cite{Ephraim1984,Ephraim1985,Martin2005a,Lotter2005,Loizou2005,Erkelens2007}.
In the well-known minimum mean-squared error (MMSE) spectral amplitude
estimator \cite{Ephraim1984}, the assumptions about the speech and
noise models are that: (a) the complex STFT coefficients of speech
and noise are additive; (b) the spectral amplitudes of speech follow
a Rayleigh distribution; (c) the additive noise is complex Gaussian
distributed. Under these assumptions, the posterior distributions
of each speech spectral amplitude has a Rician distribution whose
mean is the MMSE estimate. However, the Rayleigh assumption on the
STFT amplitudes requires the frame length to be much longer than the
correlation span within the signal. For the typical frame lengths
used in speech signal processing, this assumption is not well fulfilled
\cite{Porter1984}. Accordingly, a range of algorithms has been proposed
which assume alternative statistical distributions on either the spectral
amplitudes or the complex values of the STFT coefficients. In \cite{Martin2005a},
super-Gaussian distributions, including the Laplace and Gamma distributions,
are used to model the distribution of the real and imaginary parts
of the STFT coefficients of the speech and noise. The authors derived
MMSE estimators for when the STFT coefficients were assumed to follow
Laplacian or Gamma distributions for speech and Gaussian or Laplacian
distributions for noise. Experiments showed that estimators based
on the Laplacian speech model resulted in lower musical noise and
higher segmental SNR than the MMSE enhancers in \cite{Ephraim1984}
and \cite{Ephraim1985}. The use of the Laplacian noise model does
not lead to higher SNR values than using the Gaussian noise model
but it does result in better residual noise quality. 

Instead of an MMSE criterion, estimators can also be derived with
a maximum a posteriori (MAP) criterion \cite{Wolfe2003,Lotter2005}.
In \cite{Lotter2005}, speech spectral amplitudes are estimated using
a MAP criterion based on the Laplace and Gamma assumption on the speech
STFT coefficients. The parameters of the distributions are determined
by minimizing the Kullback-Leibler divergence against experimental
data and the noise STFT coefficients are assumed to be Gaussian distributed.
It is found that this MAP spectral amplitude estimator performs better
than the MMSE spectral amplitude estimator from \cite{Ephraim1984}
in terms of the noise attenuation especially for white noise. As a
generalization of the Gaussian and super-Gaussian prior, a generalized
Gamma speech prior was assumed in \cite{Erkelens2007} and, based
on this assumption, estimators for both the spectral amplitude and
complex STFT coefficients were derived. The MMSE amplitude estimator
derived using the generalized Gamma prior included, as special cases,
the MMSE and MAP estimators which assume Rayleigh, Laplace, and Gamma
priors, and it was found that this estimator outperformed \cite{Ephraim1984}
and gave a slightly better performance than \cite{Lotter2005} in
terms of speech distortion and noise suppression.

 \IEEEpubidadjcol
Rather than using a MAP or MMSE criterion, speech enhancers have been
proposed in which a cost function that takes into account the perceptual
characteristics of speech and noise is optimized. For example, in
\cite{Wolfe2000,Wolfe2001}, masking thresholds were incorporated
into the derivation of the optimal spectral amplitude estimators.
The threshold for each time-frequency bin was computed from a suppression
rule based on an estimate of the clean speech signal. It showed that
this estimator outperformed the MMSE estimator \cite{Ephraim1984}
and had reduced musical noise. In \cite{Loizou2005,You2005} alternative
distortion measures were used in the cost function. In \cite{You2005}
a $\beta$-order MMSE estimator was proposed where $\beta$ represented
the order of the spectral amplitude used in the calculation of the
cost function. The value of $\beta$ could also be adapted to the
SNR of each frame. The performance of this estimator was shown to
be better than both the MMSE estimator and the logMMSE estimator in
that it gave better noise reduction and better estimation of weak
speech spectral components. The estimators in \cite{Loizou2005} and
\cite{You2005} were extended in \cite{Plourde2008}, where a weighted
$\beta$-order MMSE was present. It employed a cost function which
combined the $\beta$-order compression rule and weighted Euclidean
cost function. The cost function was parameterised to model the characteristics
of the human auditory system. It was shown that the modified cost
function led to a better estimator giving consistently better performance
in both subjective and objective experiments, especially for noise
having strong high-frequency components and at low SNRs.

\subsection{Modulation Domain Speech Enhancement}

Although alternative statistical models have been extensively explored
for speech amplitude estimation, most existing estimators do not incorporate
temporal constraints on the spectral amplitudes of speech and noise
into the derivation of the estimators. The temporal dynamics of the
spectral amplitudes are characterised by the modulation spectrum and
there is evidence, both physiological and psychoacoustic, to support
the significance of the modulation domain in speech processing \cite{Drullman1994,Drullman1994a,Atlas2003,Elhilali2003,Dubbelboer2008}.
Modulation domain processing has been shown to be effective for speech
enhancement. In \cite{Hermansky1995} and \cite{Falk2007}, enhancers
were proposed using band-pass filtering of the time trajectories of
short-time power spectrum. More recently, modulation domain enhancers
\cite{Paliwal2010,So2011,Paliwal2012,Wang2013,Wang2013a,Wang2015thesis}
have been proposed that are, based on techniques conventionally applied
in the STFT domain. In \cite{Paliwal2010}, the spectral subtraction
technique was applied in the modulation domain where it outperformed
both the STFT domain spectral subtraction enhancer \cite{Boll1979}
and the MMSE enhancer \cite{Ephraim1984} in the Perceptual Evaluation
of Speech Quality (PESQ) measure \cite{Rix2001}. Similarly, an enhancer
was proposed in \cite{Paliwal2012} that applied an MMSE spectral
estimator in the modulation domain. In \cite{So2011}, a modulation-domain
Kalman filter was proposed that gave an MMSE estimate of the speech
spectral amplitudes by combining the predicted speech amplitudes with
the observed noisy speech amplitudes. It was shown that the modulation-domain
Kalman filter outperforms the time domain Kalman filter \cite{Paliwal1987}
when the enhancement performance is measured by PESQ. In \cite{So2011},
the speech and noise were assumed to be additive in the spectral amplitude
domain. Thus, there was no phase uncertainty leveraged for calculating
the MMSE estimate of the speech spectral amplitudes. Also, the speech
spectral amplitudes were assumed to be Gaussian distributed. The modulation-domain
Kalman filter enhancer in \cite{Wang2016} extended that in \cite{So2011}
from two aspects. First, the speech and noise were assumed to be additive
in the complex STFT domain. Second, the speech spectral amplitudes
were assumed to follow a form of the generalised Gamma distribution,
which was shown to be a better model than the Gaussian distribution.
The modulation-domain Kalman filter in \cite{Wang2016} only modeled
the spectral dynamics of speech, it was shown to outperform the version
of the enhancer in \cite{So2011} that also only modeled the spectral
dynamics of speech when evaluated using the PESQ and segmental SNR
(segSNR) measures \cite{Wang2016}.

\subsection{Overview of this Paper }

This paper extends the work in \cite{Wang2016} by incorporating the
spectral dynamics of both speech and noise into the modulation-domain
Kalman filter. In order to derive the MMSE estimate, we propose a
complex-valued statistical distribution denoted ``Gaussring''. This
paper is organized as follows. In Sec.~\ref{subsec:Generalized-Gamma-model},
a modulation-domain Kalman filter enhancer is described that can incorporate
one of two alternative noise models. The update step for the first
model is taken from \cite{Wang2016} and is briefly described in Sec.~\ref{subsec:Generalized-Gamma-Speech}.
The update step for the second model is based on the proposed Gaussring
distribution and is presented in Sec.~\ref{sec:Enhancement-with-Gaussring-Priors}.
Experimental results with the proposed Gaussring model based modulation-domain
Kalman filter are shown in Sec.~\ref{sec:Implementation-and-evaluation}.
Finally, in Sec.~\ref{sec:Conclusion}, conclusions are given.

\begin{figure}
\noindent \begin{centering}
\includegraphics[scale=0.35]{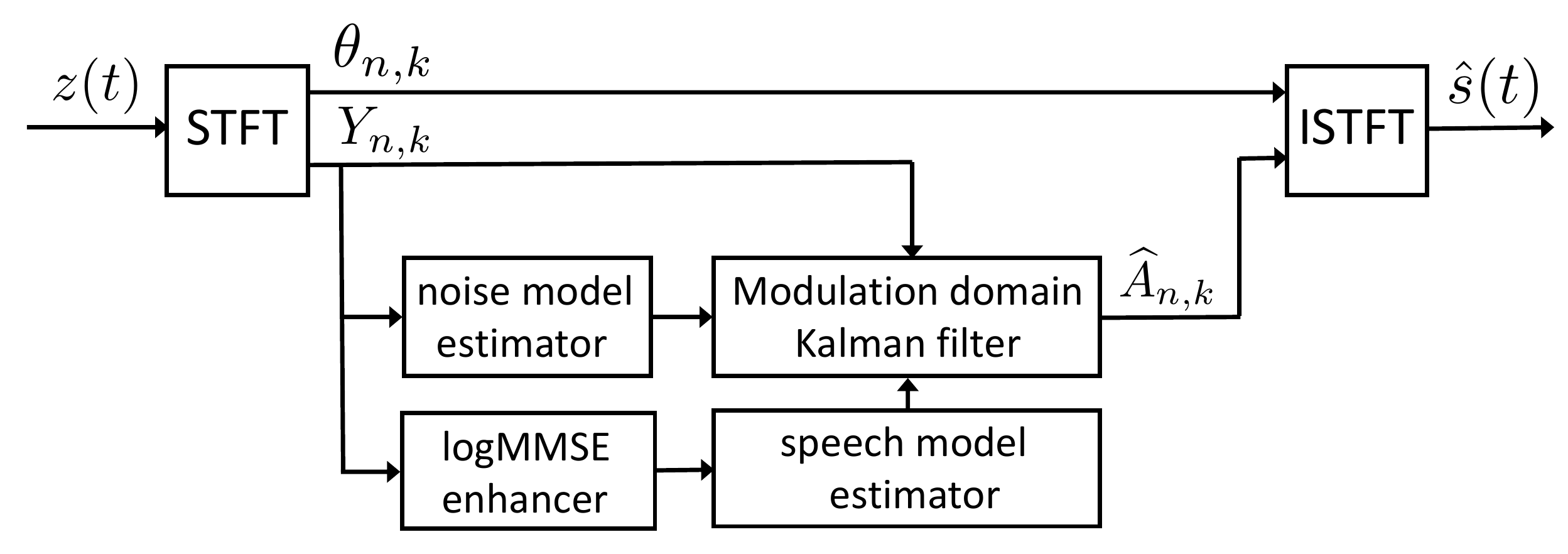}
\par\end{centering}
\caption{Diagram of proposed modulation-domain Kalman filter based MMSE estimator.
\label{fig:Diagram-of-KFMMSE-enhancer}}
\end{figure}

\section{Modulation-domain Kalman filter based MMSE enhancer\label{subsec:Generalized-Gamma-model}}

A block diagram of the modulation-domain Kalman filter based enhancement
structure is shown in Fig.~\ref{fig:Diagram-of-KFMMSE-enhancer}.
The noisy speech, $z(t)$, is transformed into the STFT domain and
enhancement is performed independently in each frequency bin, $k$.
The ``noise model estimator'' block uses the noisy speech amplitudes,
$Y_{n,k}$, where $n$ is the index for time frame, to estimate the
prior noise model. The ``speech model estimator'' block uses the
output from a logMMSE enhancer \cite{Ephraim1985,Brookes1998-2012a}
to estimate the speech model. The use of a logMMSE enhancer to pre-clean
the speech reduces the effect of the noise on the estimation of the
speech model \cite{So2011}. The modulation-domain Kalman filter combines
the speech and noise models with the observed noisy speech, $Y_{n,k}$,
to obtain an MMSE estimate of the speech spectral amplitudes, $\widehat{A}_{n,k}$.
The estimated speech is then combined with the noisy phase spectrum,
$\theta_{n,k}$, and the inverse STFT (ISTFT) is applied to obtain
the enhanced speech signal, $\hat{s}(t)$. 

\subsection{Kalman Filter Prediction Step}

The ``modulation domain Kalman filter'' block in Fig.~\ref{fig:Diagram-of-KFMMSE-enhancer}
comprises a prediction step and an update step. For frequency bin
$k$ of frame $n$, we assume that
\begin{equation}
Z_{n,k}=S_{n,k}+W_{n,k}\label{eq:complex-additive-model}
\end{equation}
where $Z_{n,k}$, $S_{n,k}$ and $W_{n,k}$ are random variables representing
the complex STFT coefficients of the noisy speech, clean speech and
noise respectively with realizations $z_{n,k}$, $s_{n,k}$ and $w_{n,k}$.
Since each frequency bin is processed independently within our algorithm,
the frequency index, $k$, will be omitted in the remainder of this
paper. The random variables representing the corresponding spectral
amplitudes are denoted: {\small{}$Y_{n}=|Z_{n}|$}, {\small{}$\tilde{A}_{n}=|S_{n}|$},
and{\small{} $\breve{A}_{n}=|W_{n}|$} with realizations $y_{n}$,
$\tilde{a}_{n}$, and $\breve{a}_{n}$. Throughout this paper, tilde,
$\sim$, and breve, $\smallsmile$, diacritics will denote quantities
relating to the estimated speech and noise signals respectively. The
prediction model assumed for the clean speech spectral amplitude is
given by

\begin{gather}
\left[\begin{array}{c}
\tilde{\mathbf{a}}_{n}\\
\breve{\mathbf{a}}_{n}
\end{array}\right]=\left[\begin{array}{cc}
\tilde{\mathbf{F}}_{n} & \mathbf{0}\\
\mathbf{0} & \mathbf{\breve{\mathbf{F}}}_{n}
\end{array}\right]\left[\begin{array}{c}
\tilde{\mathbf{a}}_{n-1}\\
\breve{\mathbf{a}}_{n-1}
\end{array}\right]+\left[\begin{array}{cc}
\tilde{\mathbf{d}} & \mathbf{0}\\
\mathbf{0} & \breve{\mathbf{d}}
\end{array}\right]\left[\begin{array}{c}
\tilde{e}_{n}\\
\breve{e}_{n}
\end{array}\right],\label{eq:state_vector_prediction}
\end{gather}
where {\small{}$\tilde{\mathbf{a}}_{n}=\left[\tilde{A}_{n},\tilde{A}_{n-1}\ldots\tilde{A}_{n-p+1}\right]^{\mathsf{T}}$}
denotes the state vector of speech amplitudes. $\tilde{\mathbf{F}}_{n}$
denotes the transition matrix for the speech amplitudes. $\tilde{\mathbf{d}}=\left[1\:0\:\cdots\:0\right]^{\mathsf{T}}$
is a $p$-dimensional vector. The speech transition matrix has the
form 

\begin{equation}
\tilde{\mathbf{F}}_{n}=\left[\begin{array}{c}
-\tilde{\mathbf{b}}_{n}^{\mathsf{T}}\\
\begin{array}{cc}
\mathbf{I} & \mathbf{0}\end{array}
\end{array}\right],\label{eq:trainsition_matrix_Chapter_5}
\end{equation}
where $\tilde{\mathbf{b}}_{n}=\left[b_{n1}\:\cdots\:b_{np}\right]^{\mathsf{T}}$
is the LPC coefficient vector, $\mathbf{I}$ is an identity matrix
of size $(p-1)\times(p-1)$ and \textbf{$\mathbf{0}$ }denotes an
all-zero column vector of length $p-1$. $\tilde{e}_{n}$ represents
the prediction residual signal and it has variance $\tilde{\eta}^{2}$.
The quantities $\breve{\mathbf{a}}_{n}$, $\breve{\mathbf{F}}_{n}$,
$\breve{\mathbf{d}}$ and $\breve{e}_{n}$ are defined similarly for
the order-$q$ noise model. By concatenating the speech and noise
state vectors, we can rewrite (\ref{eq:state_vector_prediction})
more compactly as

\begin{equation}
\mathbf{a}_{n}=\mathbf{F}_{n}\mathbf{a}_{n-1}+\mathbf{D}\mathbf{e}_{n}.\label{eq:compact_model}
\end{equation}
where the quantities, $\mathbf{a}_{n}$, $\mathbf{F}_{n}$, $\mathbf{D}$
and $\mathbf{e}_{n}$, have been defined in (\ref{eq:state_vector_prediction})
and $\mathbf{a}_{n}=\left[\begin{array}{c}
\tilde{\mathbf{a}}_{n}\end{array}\breve{\mathbf{a}}_{n}\right]^{\mathsf{T}}$, $\mathbf{F}_{n}=\left[\begin{array}{cc}
\tilde{\mathbf{F}}_{n} & \mathbf{0}\\
\mathbf{0} & \mathbf{\breve{\mathbf{F}}}_{n}
\end{array}\right]$, $\mathbf{D}=\left[\begin{array}{cc}
\tilde{\mathbf{d}} & \mathbf{0}\\
\mathbf{0} & \breve{\mathbf{d}}
\end{array}\right]$ and $\mathbf{e}_{n}=\left[\begin{array}{c}
\tilde{e}_{n}\end{array}\breve{e}_{n}\right]^{\mathsf{T}}$. The Kalman filter prediction step estimates the state vector mean
$\mathbf{a}_{n|n-1}$, and covariance, $\mathbf{P}_{n|n-1}$, at time
$n$ from their estimates, $\mathbf{a}_{n-1|n-1}$ and $\mathbf{P}_{n-1|n-1}$
at time $n-1$. The notation $n|n-1$ represents the prior estimate
at acoustic frame $n$ given the observation of all the previous frames
$1,\ldots,n-1$. The prediction model equations can be written as
\begin{align}
\mathbf{a}_{n|n-1} & =\mathbf{F}_{n}\mathbf{a}_{n-1|n-1}\\
\mathbf{P}_{n|n-1} & =\mathbf{F}_{n}\mathbf{P}_{n-1|n-1}\mathbf{F}_{n}^{\mathsf{T}}+\mathbf{D}\mathbf{Q}_{n}\mathbf{D}^{\mathsf{T}},\label{eq:covariance_prediction}
\end{align}
where $\mathbf{Q}_{n}=\left[\begin{array}{cc}
\tilde{\eta}^{2} & 0\\
0 & \breve{\eta}^{2}
\end{array}\right]$ is the covariance matrix of the prediction residual signal of speech
and noise. The values of $\mathbf{F}_{n}$ and $\mathbf{Q}_{n}$ are
determined from linear predictive (LPC) analysis on modulation frames
as described in Sec~\ref{sec:Implementation-and-evaluation}. The
prior mean and covariance matrix are given by
\begin{align}
\boldsymbol{\mu}{}_{n|n-1} & \triangleq\left[\begin{array}{cc}
\tilde{\mu}_{n|n-1} & \breve{\mu}_{n|n-1}\end{array}\right]^{\mathsf{T}}=\mathbf{D}^{\mathsf{T}}\mathbf{a}_{n|n-1}\label{eq:mean_mu}\\
\boldsymbol{\Sigma}_{n|n-1} & \triangleq\left[\begin{array}{cc}
\tilde{\sigma}_{n|n-1}^{2} & \varsigma_{n|n-1}\\
\varsigma_{n|n-1} & \breve{\sigma}_{n|n-1}^{2}
\end{array}\right]=\mathbf{D}^{\mathsf{T}}\mathbf{P}_{n|n-1}\mathbf{\mathbf{D}},\label{eq:variance_sigma}
\end{align}
where the matrix $\mathbf{D}$ has been defined in (\ref{eq:compact_model}).
$\tilde{\mu}_{n|n-1}$ and $\breve{\mu}_{n|n-1}$ denote the prior
estimate of the speech and noise spectral amplitude in the current
frame $n$. $\tilde{\mu}_{n|n-1}$ corresponds to the first element
of the state vector $\mathbf{a}_{n|n-1}$ and $\breve{\mu}_{n|n-1}$
corresponds to the $(p+1)$th elements of the state vector, $\mathbf{a}_{n|n-1}$.
$\tilde{\sigma}_{n|n-1}^{2}$ and $\breve{\sigma}_{n|n-1}^{2}$ denote
the variance of the prior estimate of the speech and noise and $\varsigma_{n|n-1}$
denotes the covariance between them. 

\subsection{Kalman Filter Update Step\label{subsec:Kalman-Filter-Update}}

For the update step, we first define a $\left(p+q\right)\times\left(p+q\right)$
permutation matrix, $\mathbf{V}$, such that $\mathbf{V}\mathbf{a}_{n|n-1}$
swaps elements $2$ and $p+1$ of the prior state vector $\mathbf{a}_{n|n-1}$
so that the first two elements now correspond to the speech and noise
amplitudes of frame $n$. The covariance matrix $\mathbf{P}_{n|n-1}$
can then be decomposed as 

\begin{equation}
\mathbf{P}_{n|n-1}=\mathbf{V}^{\mathsf{T}}\left[\begin{array}{cc}
\boldsymbol{\Sigma}_{n|n-1} & \mathbf{\mathbf{M}}_{n}^{\mathsf{\mathsf{T}}}\\
\mathbf{M}_{n} & \mathbf{T}_{n}
\end{array}\right]\mathbf{V},
\end{equation}
where $\mathbf{M}_{n}$ is a $\left(p+q-2\right)\times2$ matrix and
$\mathbf{T}_{n}$ is a $\left(p+q-2\right)\times\left(p+q-2\right)$
matrix. We now define a transformed state vector, $\mathbf{x}_{n|n-1}$
to be

\[
\mathbf{x}_{n|n-1}=\mathbf{H}_{n}\mathbf{a}_{n|n-1},
\]
where the transformation matrix is given by 

\[
\mathbf{H}_{n}=\left[\begin{array}{cc}
\mathbf{I}_{(2)} & \mathbf{0}^{\mathsf{T}}\\
-\mathbf{M}_{n}\mathbf{\boldsymbol{\Sigma}}_{n|n-1}^{-1} & \mathbf{I}_{(p+q-2)}
\end{array}\right]\mathbf{V},
\]
where $\mathbf{I}_{(j)}$ is the $j\times j$ identity matrix. 

The covariance matrix of $\mathbf{x}_{n|n-1}$ is given by 

{\small{}
\begin{align*}
\textrm{Cov}\left(\mathbf{x}_{n|n-1}\right) & =\mathbf{H}_{n}\mathbf{P}_{n|n-1}\mathbf{H}_{n}^{\intercal}\\
 & =\left[\begin{array}{cc}
\boldsymbol{\Sigma}_{n|n-1} & \mathbf{0}^{\mathsf{T}}\\
\mathbf{0} & \mathbf{T}_{n}-\mathbf{M}_{n}(\boldsymbol{\Sigma}_{n|n-1}^{-1})\mathbf{M}_{n}^{\mathsf{T}}
\end{array}\right].
\end{align*}
}It can be seen that the first two elements in the transformed state
vector are uncorrelated with other elements. Suppose the posterior
estimate of the speech and noise amplitude and the corresponding covariance
matrix in the current frame are determined to be $\boldsymbol{\mu}_{n|n}$
and $\boldsymbol{\Sigma}_{n|n}$, respectively. The state vector can
be updated as
\begin{align*}
\mathbf{x}_{n|n} & =\mathbf{x}_{n|n-1}+\mathbf{D}\left(\mathbf{\boldsymbol{\mu}}_{n|n}-\mathbf{D}^{\mathsf{T}}\mathbf{x}_{n|n-1}\right)
\end{align*}
from which, applying the inverse transformation,
\begin{equation}
\mathbf{a}_{n|n}=\mathbf{H}_{n}^{-1}\left(\mathbf{x}_{n|n-1}+\mathbf{D}\left(\mathbf{\boldsymbol{\mu}}_{n|n}-\mathbf{D}^{\mathsf{T}}\mathbf{x}_{n|n-1}\right)\right)\label{eq:state-vector-update}
\end{equation}

The covariance matrix, $\mathbf{P}_{n|n}$, can similarly be calculated
as
\begin{align}
\mathbf{P}_{n|n} & =\mathbf{H}_{n}^{-1}\left[\begin{array}{cc}
\boldsymbol{\Sigma}_{n|n} & \mathbf{0}^{\mathsf{T}}\\
\mathbf{0} & \mathbf{T}_{n}-\mathbf{M}_{n}(\mathbf{\boldsymbol{\Sigma}}_{n|n-1}^{-1})\mathbf{M}_{n}^{\mathsf{T}}
\end{array}\right]\mathbf{H}_{n}^{-\mathsf{T}}\nonumber \\
 & =\mathbf{P}_{n|n-1}+\mathbf{H}_{n}^{-1}\mathbf{D}\left(\boldsymbol{\Sigma}_{n|n}-\boldsymbol{\Sigma}_{n|n-1}\right)\mathbf{D}^{\mathsf{T}}\mathbf{H}_{n}^{-\mathsf{T}}\label{eq:covariance-matrix-update}
\end{align}

It worth noting that this formulation for the posterior estimate is
equivalent to that in \cite{Gibson1991,So2011} if the prior distribution
of the state vector is assumed to follow a Gaussian distribution but
it also allows the use of non-Gaussian distributions for the prior
estimate.

\section{Posterior distribution\label{sec:Posterior-distribution}}

\subsection{MMSE estimate}

To perform the Kalman filter update step in Sec.~\ref{subsec:Kalman-Filter-Update},
we need to obtain the posterior estimate of the state vector, $\mathbf{\boldsymbol{\mu}}_{n|n}$,
and covariance matrix, $\mathbf{\boldsymbol{\Sigma}}_{n|n}$. The
MMSE estimate of the state vector is given by the expectation of the
posterior distribution 
\begin{align}
\mathbf{\boldsymbol{\mu}}_{n|n} & =\mathbb{E}\left(\left[\tilde{A}_{n}\,\breve{A}_{n}\right]^{\mathsf{T}}|\mathcal{Y}_{n}\right)=\mathbf{D}^{\mathsf{T}}\mathbf{a}_{n|n}\nonumber \\
 & =\left[\int_{0}^{\infty}p(\tilde{a}_{n}|\mathcal{Y}_{n})d\tilde{a}_{n}\,\int_{0}^{\infty}p(\breve{a}_{n}|\mathcal{Y}_{n})d\breve{a}_{n}\right]^{\mathsf{T}},\label{eq:estimator1}
\end{align}
where $\mathcal{Y}_{n}=\left[Y_{1}\ldots Y_{n}\right]$ represents
the observed noisy speech amplitudes up to time $n$. The covariance
matrix is given by
\begin{equation}
\boldsymbol{\Sigma}_{n|n}=\mathbb{E}\left(\left[\begin{array}{cc}
\tilde{A}_{n}^{2} & \tilde{A}_{n}\breve{A}_{n}\\
\breve{A}_{n}\tilde{A}{}_{n} & \breve{A}_{n}^{2}
\end{array}\right]|\mathcal{Y}_{n}\right)-\mathbf{\boldsymbol{\mu}}_{n|n}\mathbf{\boldsymbol{\mu}}_{n|n}^{\mathsf{T}}.\label{eq:estimator1_covariance}
\end{equation}

Using Bayes rule, the posterior distribution of speech amplitudes,
$p(\tilde{a}_{n}|\mathcal{Y}_{n})$, is calculated as {\small{}
\begin{gather}
p\left(\tilde{a}_{n}|\mathcal{Y}_{n}\right)=p\left(\tilde{a}_{n}|z_{n},\mathcal{Y}_{n-1}\right)=\int_{-\pi}^{\pi}p\left(\tilde{a}_{n},\phi_{n}|z_{n},\mathcal{Y}_{n-1}\right)d\phi_{n}\nonumber \\
=\frac{\int_{-\pi}^{\pi}p(z_{n}|\tilde{a}_{n},\phi_{n},\mathcal{Y}_{n-1})p\left(\tilde{a}_{n},\phi_{n}|\mathcal{Y}_{n-1}\right)d\phi_{n}}{p\left(z_{n}|\mathcal{Y}_{n-1}\right)}\nonumber \\
=\frac{\int_{-\pi}^{\pi}p\left(w_{n}=z_{n}-\tilde{a}_{n}e^{j\phi_{n}}|\tilde{a}_{n},\phi_{n},\mathcal{Y}_{n-1}\right)p\left(\tilde{a}_{n},\phi_{n}|\mathcal{Y}_{n-1}\right)d\phi_{n}}{p\left(z_{n}|\mathcal{Y}_{n-1}\right)}\nonumber \\
=\frac{\int_{-\pi}^{\pi}p\left(w_{n}=z_{n}-\tilde{a}_{n}e^{j\phi_{n}}|\mathcal{Y}_{n-1}\right)p\left(\tilde{a}_{n},\phi_{n}|\mathcal{Y}_{n-1}\right)d\phi_{n}}{p\left(z_{n}|\mathcal{Y}_{n-1}\right)}\nonumber \\
=\frac{\int_{-\pi}^{\pi}p\left(w_{n}=z_{n}-\tilde{a}_{n}e^{j\phi_{n}}|\mathcal{Y}_{n-1}\right)p\left(\tilde{a}_{n},\phi_{n}|\mathcal{Y}_{n-1}\right)d\phi_{n}}{\int_{0}^{\infty}\int_{-\pi}^{\pi}p\left(w_{n}=z_{n}-\tilde{a}_{n}e^{j\phi_{n}}|\mathcal{Y}_{n-1}\right)p\left(\tilde{a}_{n},\phi_{n}|\mathcal{Y}_{n-1}\right)d\phi_{n}d\tilde{a}_{n}}\label{eq:conditional probability}
\end{gather}
}where $\phi_{n}$ is the realization of the random variable $\Phi_{n}$
which represents the phase of the clean speech. $p\left(z_{n}|\tilde{a}_{n},\phi_{n},\mathcal{Y}_{n-1}\right)=p\left(w_{n}=z_{n}-\tilde{a}_{n}e^{j\phi_{n}}|\tilde{a}_{n},\phi_{n},\mathcal{Y}_{n-1}\right)$
is the observation likelihood and equals the conditional distribution
of the noise, $W_{n}$. The distribution $p\left(\tilde{a}_{n},\phi_{n}|\mathcal{Y}_{n-1}\right)$
is the prior model of the speech amplitudes and its mean and variances
can be obtained from the Kalman filter prediction step given in (\ref{eq:mean_mu})
and (\ref{eq:variance_sigma}). Analogous to (\ref{eq:conditional probability}),
the posterior distribution of the noise, $p\left(\breve{a}_{n}|\mathcal{Y}_{n}\right)$,
can be calculated in a similar way. 

\begin{figure}
\begin{raggedright}
\includegraphics[scale=0.365]{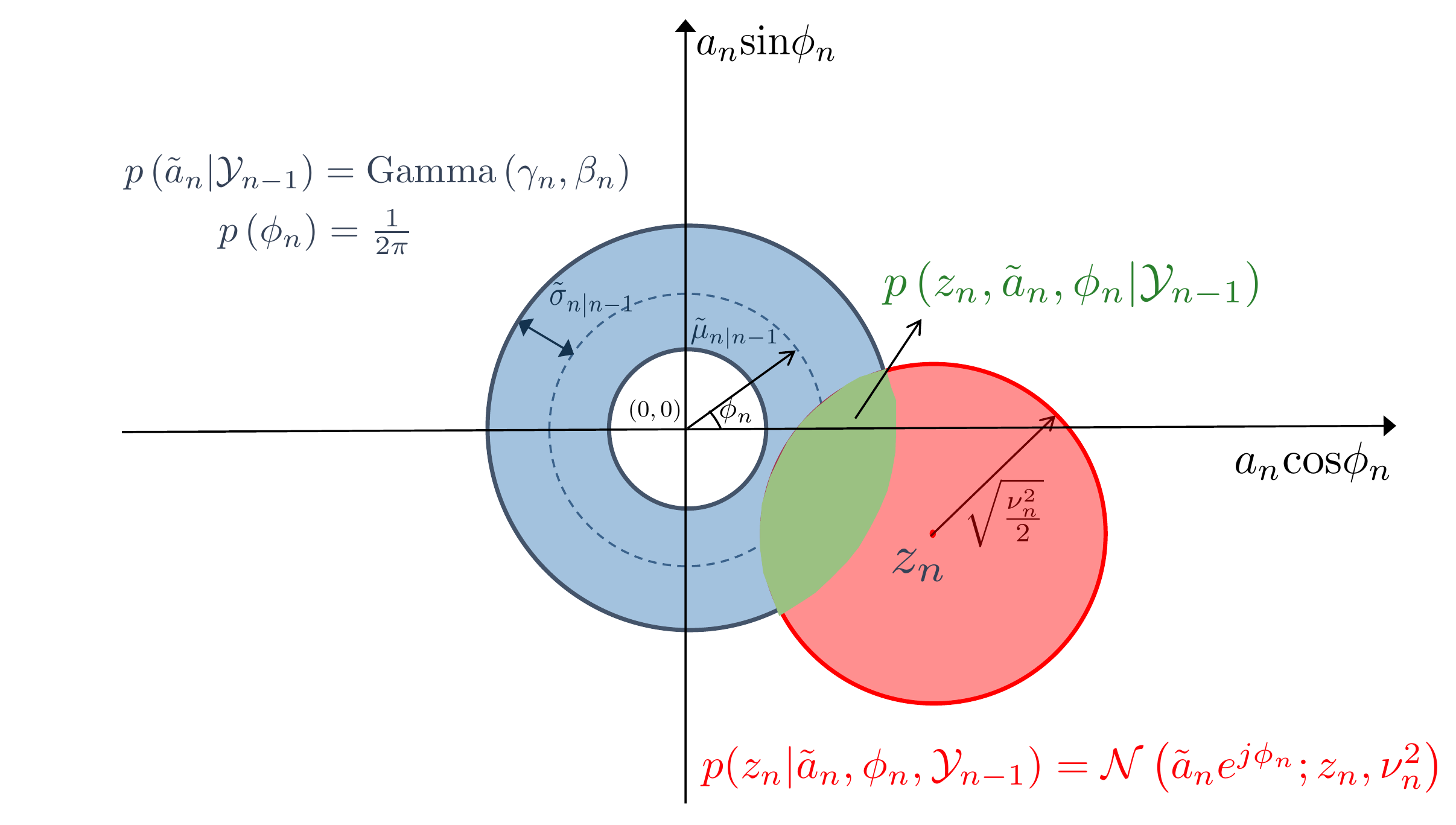}
\par\end{raggedright}
\caption{Statistical model assumed in the derivation of the posterior distribution.
The blue ring-shape distribution centered on the origin represents
the prior model: Gamma distributed in amplitude (\ref{eq:gammadist})
(denoted as $\mathrm{Gamma}$($\cdot$)) and uniform in phase. The
red circle centered on the observation, $z_{n}$, represents the Gaussian
observation likelihood model (\ref{eq:observation-likelihood}). The
green lens represents the posterior distribution, which is proportional
to the product of the other two. \label{fig:Statistical-model-KMMSE}}
\end{figure}

\subsection{Generalized Gamma Speech Prior \label{subsec:Generalized-Gamma-Speech}}

In this section, which is based on \cite{Wang2016}, the distribution
of the prior speech amplitude \textsl{$p\left(\tilde{a}_{n}|\mathcal{Y}_{n-1}\right)$}
is modeled using a 2-parameter Gamma distribution 
\begin{equation}
p\left(\tilde{a}_{n}|\mathcal{Y}_{n-1}\right)=\frac{2\tilde{a}_{n}^{2\gamma_{n}-1}}{\beta_{n}^{2\gamma_{n}}\Gamma\left(\gamma_{n}\right)}\exp\left(-\frac{\tilde{a}_{n}^{2}}{\beta_{n}^{2}}\right),\label{eq:gammadist}
\end{equation}
where $\Gamma\left(\cdot\right)$ is the Gamma function. The update
equations induced by this prior were first derived in \cite{Wang2016};
they are included here as (\ref{eq:mean-1}) and (\ref{eq:variance-1}).
The two parameters, $\beta_{n}$ and $\gamma_{n}$ are chosen to match
the mean $\mu_{n}$ and variance $\sigma_{n}^{2}$ of the predicted
amplitude given by (\ref{eq:mean_mu}) and (\ref{eq:variance_sigma}):

{\small{}
\begin{align}
\beta_{n}\frac{\Gamma\left(\gamma_{n}+0.5\right)}{\Gamma\left(\gamma_{n}\right)} & =\tilde{\mu}_{n|n-1}^{2},\label{eq:tmp1}\\
\beta_{n}^{2}\left(\gamma_{n}-\frac{\Gamma^{2}\left(\gamma_{n}+0.5\right)}{\Gamma^{2}\left(\gamma_{n}\right)}\right) & =\tilde{\sigma}_{n|n-1}^{2}.\label{eq:tmp2}
\end{align}
}{\small \par}

Eliminating $\beta_{n}$ between these equations gives
\begin{equation}
\frac{\Gamma^{2}\left(\gamma_{n}+0.5\right)}{\gamma_{n}\Gamma^{2}\left(\gamma_{n}\right)}=\frac{\tilde{\mu}_{n|n-1}^{2}}{\tilde{\mu}_{n|n-1}^{2}+\tilde{\sigma}_{n|n-1}^{2}}
\end{equation}
where $\Gamma(\cdotp)$ is the gamma function. Following \cite{Wang2016},
the solution to this equation can be approximated as $\gamma_{n}=\mathrm{tan}\left(f\left(\frac{\tilde{\mu}_{n|n-1}^{2}}{\tilde{\mu}_{n|n-1}^{2}+\tilde{\sigma}_{n|n-1}^{2}}\right)\right)$
where $f\left(\cdot\right)$ is a quartic polynomial. The observation
noise is assumed to be complex Gaussian distributed with variance
$\nu_{n}^{2}=E\left(\breve{A}_{n}^{2}\right)$ leading to the observation
model likelihood{\small{}
\begin{gather}
p\left(w_{n}=z_{n}-\tilde{a}_{n}e^{j\phi_{n}}|\mathcal{Y}_{n-1}\right)=\frac{1}{\pi\nu_{n}^{2}}\exp\left\{ -\frac{1}{\nu_{n}^{2}}|z_{n}-\tilde{a}{}_{n}e^{j\phi_{n}}|^{2}\right\} .\label{eq:observation-likelihood}
\end{gather}
}Given the assumed prior model and the observation model, the posterior
distribution of the speech amplitude in (\ref{eq:conditional probability})
is given by substituting (\ref{eq:gammadist}) and (\ref{eq:observation-likelihood})
into (\ref{eq:conditional probability})

{\small{}
\begin{multline}
p\left(\tilde{a}_{n}|\mathcal{Y}_{n}\right)=\\
\frac{\int_{0}^{2\pi}a_{n}^{2\gamma_{n}-1}\exp\left\{ -\frac{a_{n}^{2}}{\beta_{n}^{2}}-\frac{1}{\nu_{n}^{2}}|z_{n}-a{}_{n}e^{j\phi_{n}}|^{2}\right\} d\phi_{n}}{\int_{0}^{\infty}\int_{0}^{2\pi}a_{n}^{2\gamma_{n}-1}\exp\left\{ -\frac{a_{n}^{2}}{\beta_{n}^{2}}-\frac{1}{\nu_{n}^{2}}|z_{n}-a{}_{n}e^{j\phi_{n}}|^{2}\right\} d\phi_{n}da_{n}}.\label{eq:posterior-distribution}
\end{multline}
}To illustrate (\ref{eq:conditional probability}), the update model
is depicted in Fig.~\ref{fig:Statistical-model-KMMSE}. The blue
ring-shaped distribution centered on the origin represents the prior
model, $p\left(\tilde{a}_{n},\phi_{n}|\mathcal{Y}_{n-1}\right)$,
where $\mathrm{Gamma}\left(\gamma_{n},\ \beta_{n}\right)$ denotes
the Gamma distribution from (\ref{eq:gammadist}). The red circle
centered on the observation, $z_{n}$, represents the observation
model $p\left(z_{n}|\tilde{a}_{n},\phi_{n}\right)$. As in (\ref{eq:conditional probability}),
the product of the two models gives
\begin{multline}
p\left(z_{n},\tilde{a}_{n},\phi_{n}|\mathcal{Y}_{n-1}\right)=\\
p\left(\tilde{a}_{n},\phi_{n}|\mathcal{Y}_{n-1}\right)p\left(-w_{n}=\tilde{a}_{n}e^{j\phi_{n}}-z_{n}|\tilde{a}_{n},\phi_{n},\mathcal{Y}_{n-1}\right),
\end{multline}
where the second term, represented by the red circle in Fig.~\ref{fig:Statistical-model-KMMSE},
is the distribution of $-W_{n}$ but offset by the observation $z_{n}$.
The green lens-shaped region of overlap represents the product of
these distributions, $p\left(z_{n},\tilde{a}_{n},\phi_{n}|\mathcal{Y}_{n-1}\right)$.
The posterior distribution $p\left(\tilde{a}_{n}|\mathcal{Y}_{n}\right)$
is calculated by marginalising over the phase, $\phi_{n}$, in $p\left(z_{n},\tilde{a}_{n},\phi_{n}|\mathcal{Y}_{n-1}\right)$
and normalising by the integral of the green region. Substituting
(\ref{eq:posterior-distribution}) into (\ref{eq:estimator1}), a
closed-form expression can be derived for the estimator~(\ref{eq:estimator1})
using \cite[Eq. 6.643.2, 9.210.1, 9.220.2]{Jeffrey2007}

{\small{}
\begin{align}
\tilde{\mu}_{n|n} & =\int_{0}^{\infty}\tilde{a}_{n}p(\tilde{a}_{n}|\mathcal{Y}_{n})da_{n}\label{eq:estimator2}\\
 & =\frac{\int_{0}^{\infty}\int_{0}^{2\pi}\tilde{a}_{n}^{2\gamma_{n}}\exp\left\{ -\frac{\tilde{a}_{n}^{2}}{\beta_{n}^{2}}-\frac{1}{\nu_{n}^{2}}|z_{n}-\tilde{a}{}_{n}e^{j\phi_{n}}|^{2}\right\} d\phi_{n}d\tilde{a}_{n}}{\int_{0}^{\infty}\int_{0}^{2\pi}\tilde{a}_{n}^{2\gamma_{n}-1}\exp\left\{ -\frac{\tilde{a}_{n}^{2}}{\beta_{n}^{2}}-\frac{1}{\nu_{n}^{2}}|z_{n}-\tilde{a}{}_{n}e^{j\phi_{n}}|^{2}\right\} d\phi_{n}d\tilde{a}_{n}}\nonumber \\
 & =\frac{\Gamma\left(\gamma_{n}+0.5\right)}{\Gamma\left(\gamma_{n}\right)}\sqrt{\frac{\xi_{n}}{\zeta_{n}(\gamma_{n}+\xi_{n})}}\frac{\mathcal{M}\left(\gamma_{n}+0.5;1;\frac{\zeta_{n}\xi_{n}}{\gamma_{n}+\xi_{n}}\right)}{\mathcal{M}\left(\gamma_{n};1;\frac{\zeta_{n}\xi_{n}}{\gamma_{n}+\xi_{n}}\right)}y_{n},\label{eq:mean-1}
\end{align}
}where $\mathcal{M}$ is the confluent hypergeometric function \cite{NIST2010},
and $\xi_{n}$ and $\zeta_{n}$ are the a priori SNR and a posteriori
SNR respectively, which are calculated as
\[
\zeta_{n}=\frac{y_{n}^{2}}{\nu_{n}^{2}},~~\xi_{n}=\frac{\text{\ensuremath{\mathbb{E}}}\left(\tilde{A}_{n}^{2}|\mathcal{Y}_{n-1}\right)}{\nu_{n}^{2}}=\frac{\tilde{\mu}_{n|n-1}^{2}+\tilde{\sigma}_{n|n-1}^{2}}{\nu_{n}^{2}}=\frac{\gamma_{n}\beta_{n}^{2}}{\nu_{n}^{2}}.
\]
The variance associated with the estimator in (\ref{eq:mean-1}) is
given by \cite[Eq. 6.643.2, 9.210.1, 9.220.2]{Jeffrey2007}{\small{}
\begin{align}
\tilde{\sigma}_{n|n}^{2} & =\mathbb{E}\left(\tilde{A}_{n}^{2}|\mathcal{Y}_{n},\phi_{n}\right)-\left(\mathbb{E}\left(\tilde{A}_{n}|\mathcal{Y}_{n},\phi_{n}\right)\right)^{2}\nonumber \\
 & =\frac{\gamma_{n}\xi_{n}}{\zeta_{n}(\gamma_{n}+\xi_{n})}\frac{\mathcal{M}\left(\gamma_{n}+1;1;\frac{\zeta_{n}\xi_{n}}{\gamma_{n}+\xi_{n}}\right)}{\mathcal{M}\left(\gamma_{n};1;\frac{\zeta_{n}\xi_{n}}{\gamma_{n}+\xi_{n}}\right)}y_{n}^{2}-\tilde{\mu}_{n|n-1}^{2}.\label{eq:variance-1}
\end{align}
}{\small \par}

Since the noise is assumed to be stationary and the LPC order $q=0$,
the state vector is updated in (\ref{eq:state-vector-update}) with
$\mathbf{D}=\tilde{\mathbf{d}}$ and $\boldsymbol{\mu}_{n|n}=\tilde{\mu}_{n|n}$
and the covariance matrix is updated in (\ref{eq:covariance-matrix-update})
with $\boldsymbol{\Sigma}_{n|n}=\tilde{\sigma}_{n|n}^{2}$ . 

\subsection{Enhancement with Gaussring priors\label{sec:Enhancement-with-Gaussring-Priors}}

In this section, we jointly model the temporal dynamics of spectral
amplitudes of both the speech and noise. In this case, the observation
model assumed in \cite{So2010a}, $R_{n}=A_{n}+V_{n}$, can be viewed
as a constraint applied to the speech and noise when deriving the
MMSE estimate for their amplitudes. As in Sec.~\ref{subsec:Generalized-Gamma-model},
we assume that the speech and noise are additive in the complex STFT
domain. The STFT coefficients of speech and noise are assumed to have
uniform prior phase distributions. To derive the Kalman filter update,
the joint posterior distribution of the speech and noise amplitudes
need to be estimated to apply in (\ref{eq:state-vector-update}) and
(\ref{eq:covariance-matrix-update}). However, in this case the normalisation
term in (\ref{eq:conditional probability}) is now calculated as 

{\small{}
\begin{multline}
p\left(z_{n}|\mathcal{Y}_{n-1},\mathcal{\breve{A}}_{n-1}\right)=\int_{0}^{\infty}\int_{0}^{2\pi}\int_{0}^{\infty}\int_{0}^{2\pi}p\left(z_{n}|\tilde{a}_{n},\phi_{n},\breve{a}{}_{n},\psi_{n}\right)\\
p\left(\tilde{a}_{n},\phi_{n},\breve{a}{}_{n},\psi_{n}|\mathcal{Y}_{n-1},\mathcal{V}_{n-1}\right)d\tilde{a}_{n}d\phi_{n}d\breve{a}{}_{n}d\psi{}_{n},
\end{multline}
}where {\small{}$\mathcal{\breve{A}}_{n}=\left[\breve{A}_{1}\ldots\breve{A}_{n}\right]$}
represents the noise amplitudes up to time $n$ and $\psi_{n}$ is
the realization of the random variable $\Psi_{n}$ which represents
the phase of the noise. This marginalisation is mathematically intractable
if the generalized Gamma distribution from (\ref{eq:gammadist}) is
assumed for both the speech and noise prior amplitude distributions.

In order to overcome this problem, in this section we assume the complex
STFT coefficients to follow a ``Gaussring'' distribution that comprises
a mixture of Gaussians whose centres lie in a circle on the complex
plane. 

\subsubsection{Gaussring distribution}

From the colored noise modulation-domain Kalman filter described in
\cite{So2011}, the prior estimate of the amplitude of both speech
and noise can be obtained. The idea of the Gaussring model is, to
use a mixture of $2$-dimensional circular Gaussians to approximate
the prior distribution of the complex STFT coefficients of both the
speech, $p\left(s_{n|n-1}\right)$, and the noise, $p\left(w_{n|n-1}\right)$. 

For the speech coefficients, the Gaussring model is defined as 
\begin{align}
p\left(s_{n|n-1}\right)=\sum_{\gringjs=1}^{\tilde{G}}\tilde{\mixweight}_{n|n-1}^{(\tilde{g})}\mathcal{N}\left(\tilde{\mixmean}_{n|n-1}^{\left(\gringjs\right)},\,\tilde{\mixvar}_{n|n-1}\right)\label{eq:noise_mixture}
\end{align}
where $\widetilde{G}$ is the number of Gaussian components and $\tilde{\mixweight}_{n|n-1}^{(\tilde{g})}$
is the weight of the $\tilde{g}$th Gaussian component. $\tilde{\mixmean}_{n|n-1}^{\left(\gringjs\right)}$
denotes the complex mean of the $\gringjs$th Gaussian component and
$\tilde{\mixvar}_{n|n-1}$ denotes real-valued variance (which is
common to all components). The noise Gaussring model $p\left(w_{n|n-1}\right)$
is similarly defined with parameters $\breve{G},$ $\breve{\epsilon}_{n|n-1}^{(\breve{g})}$,
$\breve{o}_{n|n-1}^{\left(\breve{g}\right)}$ and $\breve{\mixvar}_{n|n-1}$. 

In this paper, we assume that the phase distribution is uniform and
hence that all mixtures have equal weights of $\epsilon_{n|n-1}^{(g)}=\frac{1}{G}$.
We note however, that the Gaussring model can be extended to incorporate
a prior phase distribution by using unequal weights for the mixtures.
In order to fit the ring distribution to the moments of the amplitude
prior from (\ref{eq:mean_mu}) and (\ref{eq:variance_sigma}), $\mu_{n|n-1}$
and $\sigma_{n|n-1}$, the number of Gaussian components, $G$, is
chosen so that the mixture centres are separated by $2\sigma_{n|n-1}$
around a circle of radius $\mu$ in the complex plane. Accordingly,
$G$ is set to be 
\begin{equation}
G=\left\lceil \frac{\pi\mu_{n|n-1}}{\sigma_{n|n-1}}\right\rceil \label{eq:number_Gaussians}
\end{equation}
where $\left\lceil \cdotp\right\rceil $ is the ceiling function. 

\begin{figure}
\begin{centering}
{\small{}(a)}
\par\end{centering}{\small \par}
\noindent \begin{centering}
\includegraphics[scale=0.28]{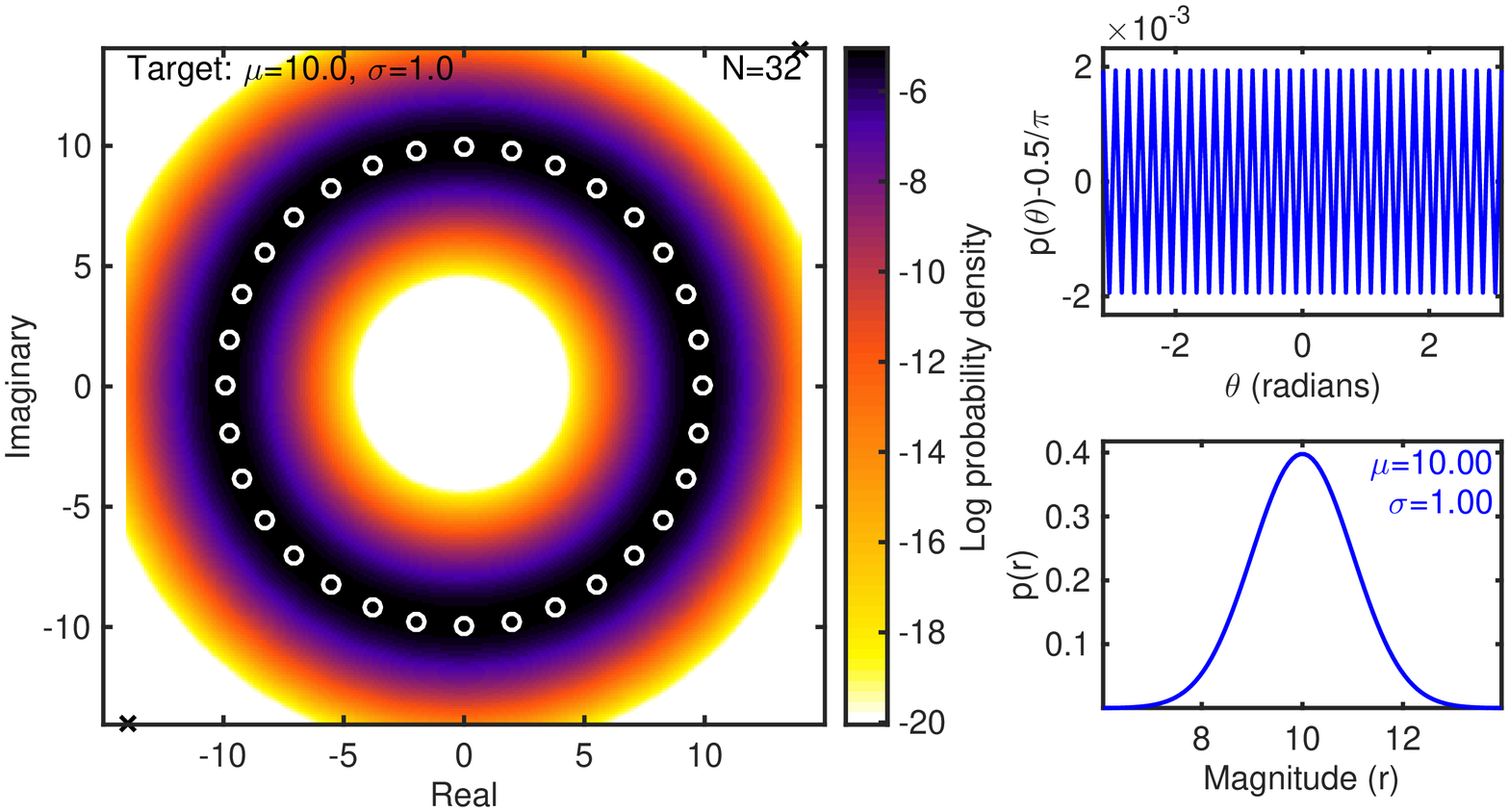}
\par\end{centering}
\begin{centering}
{\small{}(b)}
\par\end{centering}{\small \par}
\noindent \begin{centering}
\includegraphics[scale=0.28]{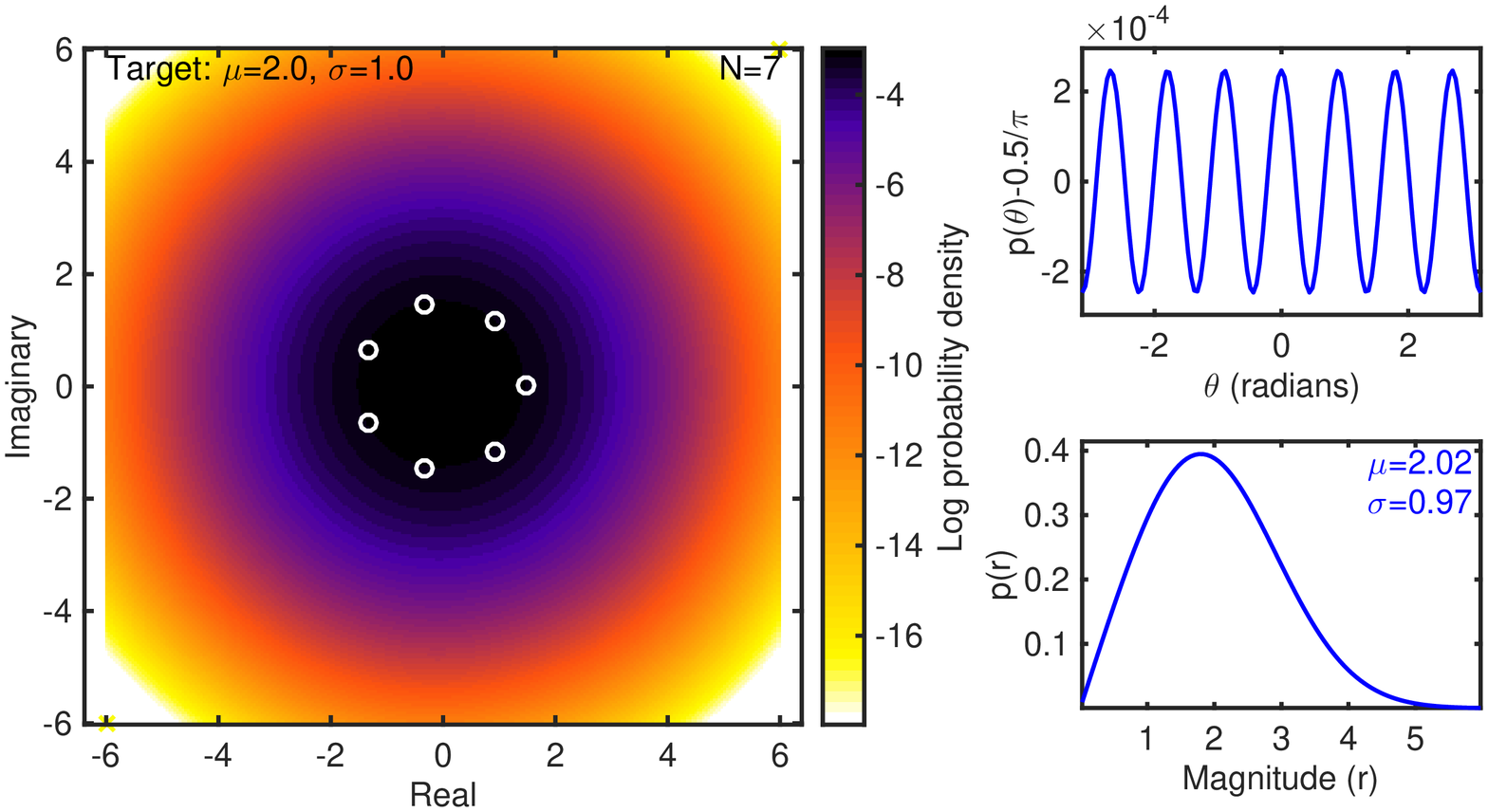}
\par\end{centering}
\begin{centering}
{\small{}(c)}
\par\end{centering}{\small \par}
\noindent \begin{centering}
\includegraphics[scale=0.28]{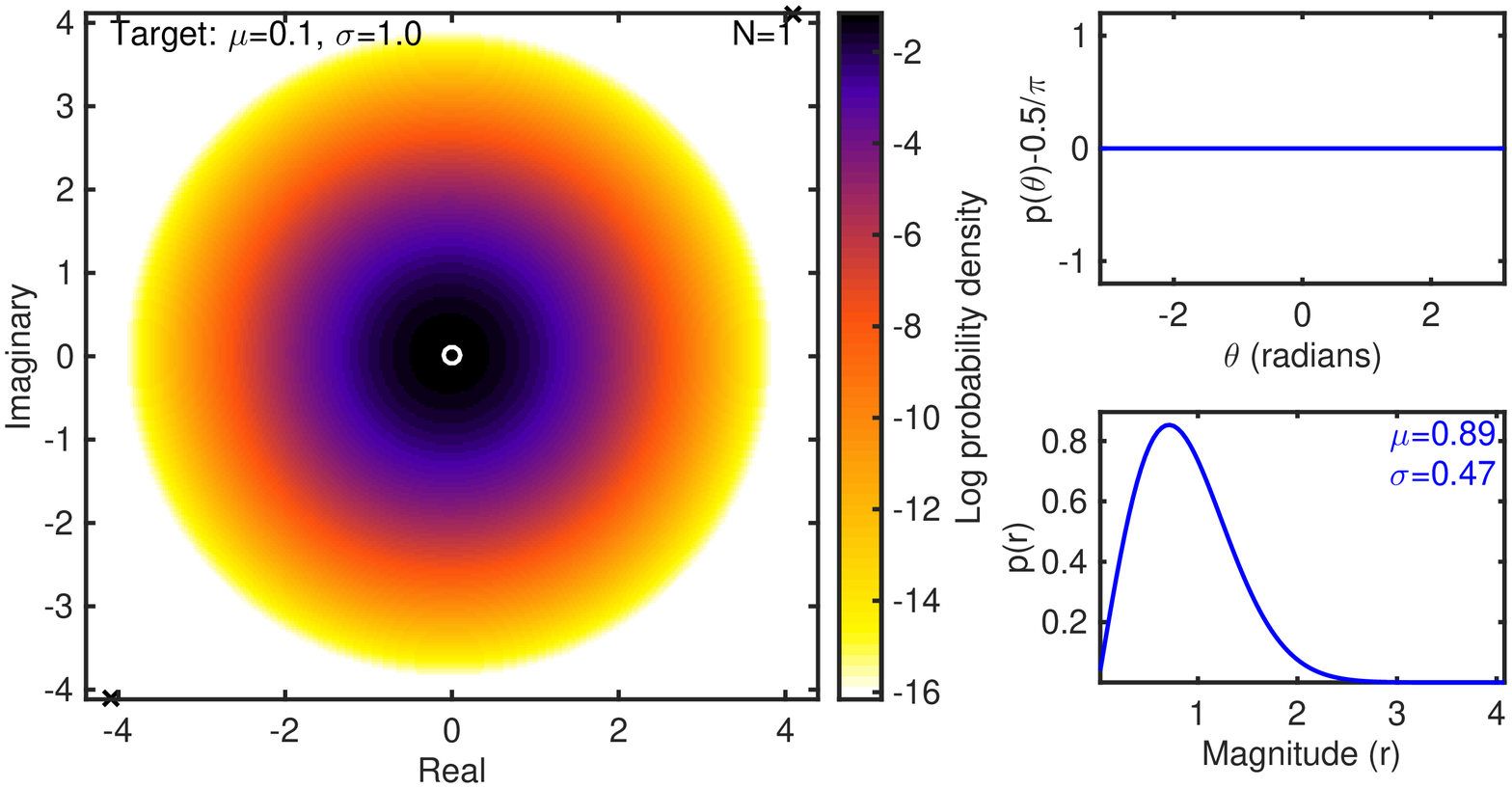}
\par\end{centering}
\caption{Gaussring model fit for targets of (a) $\mu_{n|n-1}=10.0$ and $\sigma_{n|n-1}=1.0$,
(b) $\mu_{n|n-1}=2.0$ and $\sigma_{n|n-1}=1.0$ and (c) $\mu_{n|n-1}=0.1$
and $\sigma_{n|n-1}=1.0$. The left plot shows the Gaussring distribution
in the complex plane. The two plots on the right of the figure show
the marginal distributions of phase (upper plot) and magnitude (lower
plot). \label{fig:Gaussring-Example2}}
\end{figure}

Examples of Gaussring models matching a prior estimate are shown in
Fig.~\ref{fig:Gaussring-Example2}. The left plot of Fig.~\ref{fig:Gaussring-Example2}(a)
shows the Gaussring distribution in the complex plane for the case
$\left(\mu_{n|n-1},\,\sigma_{n|n-1}\right)=\left(10,\,1\right)$ for
which $G=32$. The white circles indicate the means of the individual
Gaussian components. The two plots on the right of the figure show
the marginal distributions of phase (upper plot) and magnitude (lower
plot). The phase distribution is uniform to within $+-0.002$ and
the magnitude distribution is almost symmetric with the correct target
mean and standard deviation (printed above the plotted distribution).
Fig.~\ref{fig:Gaussring-Example2}(b) shows the same plots for the
case $\left(\mu_{n|n-1},\,\sigma_{n|n-1}\right)=\left(2,\,1\right)$
for which $G=9$. In this case the phase distribution is again close
to uniform while the amplitude distribution has almost the correct
target mean and standard deviation but is now noticeably asymmetric.
For a Rician distribution, the mean $\mu_{\mathsf{Rician}}$ and standard
deviation $\sigma_{\mathsf{Rician}}$ satisfy 
\begin{equation}
\frac{\mu_{\mathsf{Rician}}}{\sigma_{\mathsf{Rician}}}\geq\sqrt{\frac{\pi}{4-\pi}}\approx1.91\label{eq:inequality-constraint}
\end{equation}
and when $\frac{\mu_{\mathsf{Rician}}}{\sigma_{\mathsf{Rician}}}=\sqrt{\frac{\pi}{4-\pi}}$,
it becomes a Rayleigh distribution. Fig.~\ref{fig:Gaussring-Example2}(c)
illustrates the case when the target $\left(\mu_{n|n-1},\,\sigma_{n|n-1}\right)=\left(0.1,\,1\right)$
violates this condition. In this case, the model defaults to a Rayleigh
distribution whose mean square amplitude, $\mu_{n|n-1}^{2}+\sigma_{n|n-1}^{2}$
matches that of the target.

\begin{figure}
\noindent \begin{centering}
\includegraphics[scale=0.41]{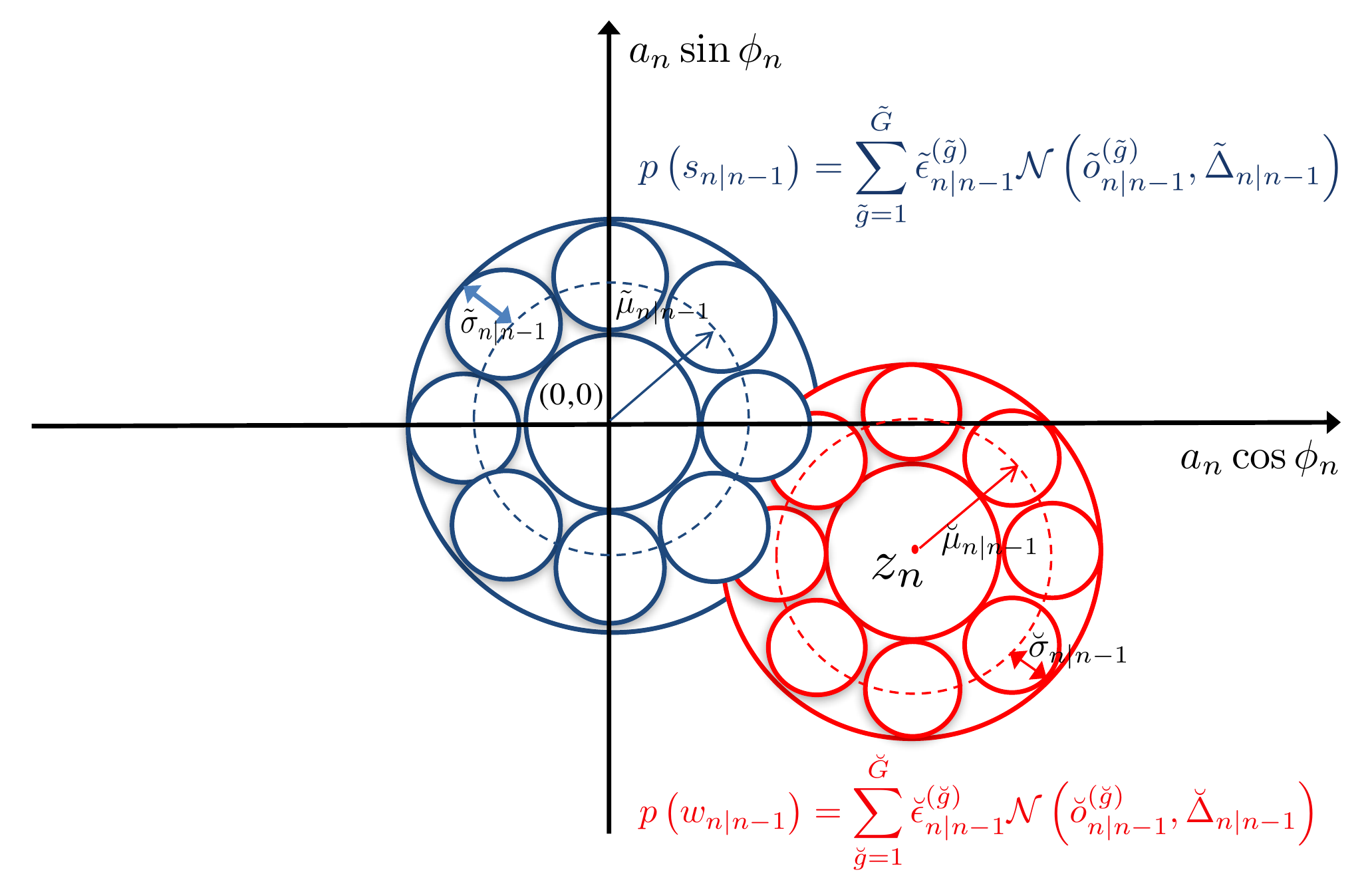}
\par\end{centering}
\caption{Gaussring model of speech and noise. Blue circles represent the speech
Guassring model and red circles represent the noise Guassring model.
\label{fig:Gaussring-model}}
\end{figure}

A diagram, analogous to Fig.~\ref{fig:Gaussring-Example2}, illustrating
a Gaussring model used for both the speech and noise priors in (\ref{eq:conditional probability})
is illustrated in Fig.~\ref{fig:Gaussring-model}. As in Fig.~\ref{fig:Gaussring-model},
the speech distribution is centered on the origin while the negated
noise distribution is centered at the observation $z_{n}$. 

Supposing that there are $\tilde{G}$ components for the speech and
$\breve{G}$ Gaussian components for the noise, a total of $\tilde{G}\breve{G}$
Gaussian components will be obtained for the posterior distribution
after combining the speech and noise prior models. The weighted product
of component of speech and component of noise, is $\mixweight_{n|n}^{\left(\gringjs,\,\gringjn\right)}\mathcal{N}\left(o_{n|n}^{\left(\gringjs,\,\gringjn\right)},\mixvar_{n|n}\right)$,
is with parameters \cite{Brookes1998-2013}

{\small{}
\begin{align}
\mixvar_{n|n} & =\frac{\tilde{\mixvar}_{n|n-1}\breve{\mixvar}_{n|n-1}}{\tilde{\mixvar}_{n|n-1}+\breve{\mixvar}_{n|n-1}}\label{eq:covariance-posterior}\\
o_{n|n}^{\left(\gringjs,\,\gringjn\right)} & =\mixvar_{n|n}\left(\frac{\tilde{\mixmean}_{n|n-1}^{(\gringjs)}}{\tilde{\mixvar}_{n|n-1}}+\frac{\breve{o}_{n|n-1}^{\left(\gringjn\right)}}{\breve{\mixvar}_{n|n-1}}\right)\label{eq:mean-posterior}\\
\mixweight_{n|n}^{\left(\gringjs,\,\gringjn\right)} & =\frac{1}{\tilde{G}\breve{G}}\mathcal{N}\left(0;\,\tilde{\mixmean}_{n|n-1}^{(\gringjs)}-\breve{o}_{n|n-1}^{\left(\gringjn\right)},\,\tilde{\mixvar}_{n|n-1}+\breve{\mixvar}_{n|n-1}\right),\label{eq:weights-posterior}
\end{align}
}where $\mathcal{N}\left(x;o,\Delta\right)$ denotes the value of
the Gaussian distribution $\mathcal{N}\left(o,\Delta\right)$ evaluated
at $x$. The optimal estimate of the amplitude of speech and noise
is calculated as the mean of the amplitude of posterior Gaussian components
as in (\ref{eq:estimator1}). 

\subsubsection{Moment Matching}

In this subsection, we will describe how the parameters of the Gaussring
model are estimated by matching the moments of the prior estimate.
Because each mixture component in the Gaussring model is circular
Gaussian, its amplitude is Rician distributed \cite{NIST2010}; with
a $2$-parameter distribution given by 
\begin{equation}
p\left(a_{n}|\mathcal{Y}_{n-1}\right)=\frac{a_{n}}{\delta^{2}}\text{exp}\left(\frac{-(a_{n}^{2}+\alpha^{2})}{2\delta^{2}}\right)I_{0}\left(\frac{a_{n}\alpha}{\delta^{2}}\right),\label{eq:Rician-distribution}
\end{equation}
where $I_{k}\left(\cdot\right)$ is a modified Bessel function of
the first kind and $a_{n}$ represents the realization of the speech
amplitude, $\tilde{a}_{n}$, or noise amplitude, $\breve{a}_{n}$.
The parameters of the Rician distribution are determined by matching
the mean and variance to $\mu_{n|n-1}$ and $\sigma_{n|n-1}$ from
(\ref{eq:mean_mu}), (\ref{eq:variance_sigma}). The mean and variance
of the Rician distribution in (\ref{eq:Rician-distribution}) are
given by
\begin{align}
\mu_{\mathsf{Rician}} & =\delta_{n}\sqrt{\frac{\pi}{2}}\text{exp}\left(-\frac{\alpha_{n}^{2}}{2\delta_{n}^{2}}\right)\nonumber \\
 & \left[\left(1-\frac{\alpha_{n}^{2}}{2\delta_{n}^{2}}\right)I_{0}\left(-\frac{\alpha_{n}^{2}}{4\delta_{n}^{2}}\right)-\frac{\alpha_{n}^{2}}{2\delta_{n}^{2}}I_{1}\left(-\frac{\alpha_{n}^{2}}{4\delta_{n}^{2}}\right)\right]\label{eq:mean-Rician}\\
\sigma_{\mathsf{Rician}}^{2} & =2\delta_{n}^{2}+\alpha_{n}^{2}-\mu_{\mathsf{Rician}}^{2},\label{eq:var-Rician}
\end{align}
where $\alpha_{n}\geq0$ and $\delta_{n}\geq0$ are the parameters
of the Rician distribution in (\ref{eq:Rician-distribution}). It
is difficult to invert (\ref{eq:mean-Rician}) to determine $\alpha$
and $\delta$ from $\mu_{n|n-1}$ and $\sigma_{n|n-1}^{2}$, so instead
we use the Nakagami-m distribution to approximate the Rician distribution.
There are two advantages to using this approximation. First, the parameters
of the distribution can be estimated efficiently by matching the moments
of the prior estimate and second, the covariance of the amplitudes
of the speech and noise can be approximated efficiently. In \cite{Xie2014},
the Nakagami-m distribution is similarly used to approximate the Rician
distribution in order to simplify the MMSE estimator in \cite{Ephraim1984}
and MAP estimator in \cite{Wolfe2003}. 

The Nakagami-m distribution is a $2$-parameter distribution given
by \cite{Cheng2001}
\[
p\left(a_{n}|\mathcal{Y}_{n-1}\right)=\frac{2m^{m}}{\Gamma\left(m\right)\Omega^{m}}a_{n}^{2m-1}\text{exp}\left(-\frac{m}{\Omega}a_{n}\right).
\]
The mean and variance of the Nakagami-m distribution are given by
\begin{align}
\mu_{\mathrm{\mathsf{Nakagami}}} & =\frac{\Gamma(m+\frac{1}{2})}{\Gamma(m)}\sqrt{\frac{\Omega}{m}}\label{eq:mean-nakagami}\\
\sigma_{\mathsf{\mathsf{Nakagami}}}^{2} & =\Omega-\mu_{\mathrm{\mathsf{Nakagami}}}^{2},\label{eq:variance-nakagami}
\end{align}
where $\Omega_{n}$ and $m_{n}$ are the parameters of the distribution
which satisfy \cite{Cheng2001}
\begin{align}
\Omega_{n} & =\text{E}\left(A_{n}^{2}\right)\label{eq:omega-nakagami}\\
m_{n} & =\frac{\text{E}^{2}\left(A_{n}^{2}\right)}{\text{Var}\left(A_{n}^{2}\right)}.\label{eq:m_nakagami}
\end{align}

The Nakagami-m distribution is a good approximation to the Rician
distribution when the parameter, $m$, in the Nakagami-m distribution
satisfies $m>1$ \cite{Wang1998,Xie2014,Crepeau1992}. The parameters
of the Rician distribution can be obtained from the parameters of
the corresponding Nakagami-m distribution for $m>1$ by moment matching
\cite{Crepeau1992} to obtain
\begin{align}
\alpha^{2} & =\Omega\sqrt{1-\frac{1}{m}}\label{eq:alpha-Rician}\\
\delta^{2} & =0.5\left(\Omega-\alpha^{2}\right).\label{eq:epslion-Rician}
\end{align}

\begin{figure}
\noindent \begin{centering}
\includegraphics[scale=0.36]{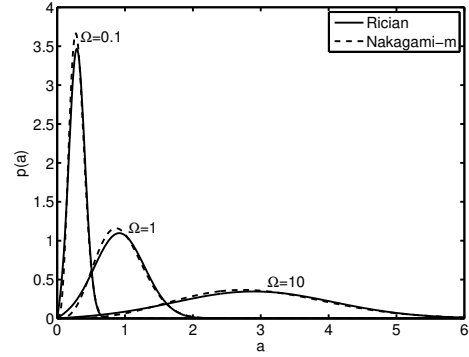}
\par\end{centering}
\caption{Comparison of Rician and Nakagami-m distribution for $\Omega=0.1,1,10$
and $m=2$.\label{fig:Comparison-of-Rician-Naka}}
\end{figure}

In Fig.~\ref{fig:Comparison-of-Rician-Naka}, the Rician distribution
and Nakamai-m distribution are compared for $\Omega=0.1,1,10$ and
$m=2$, and the parameters of Rician distribution, $\alpha$ and $\upsilon$
are calculated from $\Omega$ and $m$ using (\ref{eq:alpha-Rician})
and (\ref{eq:epslion-Rician}). It can be seen that, the Nakagami-m
distribution is a close approximation of the Rician distribution for
this range of parameters. 

It is still not straightforward to invert (\ref{eq:mean-nakagami}),
(\ref{eq:variance-nakagami}) to determine $\left(m,\,\Omega\right)$
from $\left(\mu_{n|n-1},\,\sigma_{n|n-1}^{2}\right)$. However, by
observing that $\frac{\Gamma(m+\frac{1}{2})}{\Gamma(m)}$ is tightly
bounded by \cite{Xie2014}

\begin{equation}
\sqrt{m-\frac{1}{4}}<\frac{\Gamma(m+\frac{1}{2})}{\Gamma(m)}<m/\sqrt{m+\frac{1}{4}},\label{eq:range_gamma}
\end{equation}
we can replace this quantity by its lower bound to obtain
\begin{align*}
\mu_{n|n-1} & =\sqrt{\Omega_{n|n-1}-\frac{\Omega_{n|n-1}}{4m_{n|n-1}}}\\
\sigma_{n|n-1}^{2} & =\frac{\Omega_{n|n-1}}{4m_{n|n-1}},
\end{align*}
from which
\begin{align}
\Omega_{n|n-1} & =\mu_{n|n-1}^{2}+\sigma_{n|n-1}^{2}\label{eq:omega_naka}\\
m_{n|n-1} & =0.25\sigma_{n|n-1}^{-2}\Omega_{n|n-1}.\label{eq:m-naka}
\end{align}

The $\alpha$ and $\delta^{2}$ parameters of the corresponding Rician
distribution can then be calculated from $\Omega_{n|n-1}$ and $m_{n|n-1}$
using (\ref{eq:alpha-Rician}) and (\ref{eq:epslion-Rician}). From
$\alpha$ and $\delta^{2},$ the mean and covariance of each mixture
of the Gaussring model can be obtained as
\begin{align}
\tilde{\mixmean}_{n|n-1}^{\left(\gringjs\right)} & =\tilde{\alpha}\exp\left(\frac{j2\pi\gringjs}{\tilde{G}}\right)\label{eq:mean-real}\\
\breve{o}_{n|n-1}^{\left(\gringjn\right)} & =z_{n}+\breve{\alpha}\exp\left(\frac{j2\pi\gringjn}{\breve{G}}\right)\label{eq:mean-imag}\\
\tilde{\mixvar} & =2\tilde{\delta}^{2}\quad\breve{\mixvar}=2\breve{\delta}^{2}.\label{eq:variance-real-imag}
\end{align}

When the inequality in (\ref{eq:inequality-constraint}) is not satisfied,
we use a single Gaussian component to model the distribution in (\ref{eq:noise_mixture}).
In this case, the prior distribution of the amplitude, $p\left(a_{n}|\mathcal{Y}_{n-1}\right)$,
becomes a Rayleigh distribution which is a $1$-parameter distribution.
Rather than matching the mean or variance of this Rayleigh distribution
to the corresponding prior, we estimate the parameter of the Rayleigh
distribution by matching $\text{E}\left(A_{n}^{2}|\mathcal{Y}_{n-1}\right)$,
which is calculated in (\ref{eq:omega_naka}) as $\Omega_{n|n-1}$.
Thus, the mean and variance of this Gaussian distribution is given
by $o_{n|n-1}=0$ and $\mixvar_{n|n-1}=\delta^{2}=\frac{\Omega_{n|n-1}^{2}}{2}$.
The plot in Fig.~\ref{fig:Gaussring-Example2}(c) shows the Gaussring
model with a target $\left(\mu_{n|n-1},\sigma_{n|n-1}\right)=\left(0.1,1\right)$.
We can see that the actual fitted mean and standard deviation deviate
from the actual values and are $\left(0.89,\,0.47\right)$. In this
case, the model will be fitted with a mean and standard deviation
which satisfy equality in (\ref{eq:inequality-constraint}) and give
the correct value of $\mu_{n|n-1}^{2}+\sigma_{n|n-1}^{2}$.

\subsubsection{Posterior estimate}

In order to determine the mean, $\boldsymbol{\mu}_{n|n}$, and covariance,
$\boldsymbol{\Sigma}_{n|n}$, of the posterior amplitude distribution
in (\ref{eq:estimator1}), (\ref{eq:estimator1_covariance}), we first
calculate the corresponding quantities for each Gaussian component
of the product, $\mathcal{N}\left(o_{n|n}^{\left(\gringjs,\,\gringjn\right)},\mixvar_{n|n}\right)$
from (\ref{eq:covariance-posterior}), (\ref{eq:mean-posterior}).
We use the Nakagami-m distribution to model the amplitude distribution
of this complex Gaussian, $p\left(a_{n}^{\left(\tilde{g},\breve{g}\right)}|\mathcal{Y}_{n}\right)$.
The Nakagami-m parameters, $m_{n|n}^{\left(\gringjs,\,\gringjn\right)}$
and $\Omega_{n|n}^{\left(\gringjs,\,\gringjn\right)}$, are calculated
in (\ref{eq:omega-nakagami}) and (\ref{eq:m_nakagami}) from the
mean and variance of the squared amplitude, denoted here by {\small{}$\mu_{\mathsf{sq}}^{\left(\tilde{g},\breve{g}\right)}=\text{E}\left(A_{n}^{2\left(\tilde{g},\,\breve{g}\right)}|\mathcal{Y}_{n}\right)$}
and {\small{}$\sigma_{\mathsf{sq}}^{2\left(\tilde{g},\breve{g}\right)}=\text{Var}\left(A_{n}^{2\left(\tilde{g},\,\breve{g}\right)}|\mathcal{Y}_{n}\right)$}
respectively. 

We define a 2-element complex Gaussian vector $\boldsymbol{\upsilon}\backsim\mathcal{N}\left(\meanjj,\varjj\right)$
in which the two elements are fully correlated with each other and
differ only in their means. The mean and the covariance matrix of
this vector is given by
\begin{align*}
\meanjj & =\left[\begin{array}{c}
o_{n|n}^{\left(\gringjs,\,\gringjn\right)}\end{array},~o_{n|n}^{\left(\gringjs,\,\gringjn\right)}-z_{n}\right]^{\mathsf{T}}\\
\varjj & =\left[\begin{array}{cc}
\mixvar_{n|n} & \mixvar_{n|n}\\
\mixvar_{n|n} & \mixvar_{n|n}
\end{array}\right]
\end{align*}
 from \cite{Miller1974,Brookes1998-2013} we can obtain

{\small{}
\begin{alignat}{1}
\meansqjj & =\mathrm{diag}\left(\mathrm{\boldsymbol{\Sigma}}^{\left(\tilde{g},\,\breve{g}\right)}\right)+\left|\meanjj\right|^{\circ2}\label{eq::expectation-of-square-of-amplitude}\\
\varsqjj & =\left|\varjj+\meanjj\left(\meanjj\right)^{\mathsf{H}}\right|^{\circ2}-\left|\meanjj\left(\meanjj\right)^{\mathsf{H}}\right|^{\circ2},\label{eq:variance-of-square-of-amplitude}
\end{alignat}
}in which $^{\circ2}$ and $\left|\cdot\right|$ denote element-wise
squaring and absolute value of matrix elements. These quantities may
be decomposed as

\begin{equation}
\meansqjj=\left[\tilde{\mu}_{\mathsf{sq}}^{\left(\tilde{g},\,\breve{g}\right)},\ \breve{\mu}_{\mathsf{sq}}^{\left(\tilde{g},\,\breve{g}\right)}\right]^{\mathsf{T}}\label{eq:mean_vector_sq}
\end{equation}
and

\begin{equation}
\varsqjj=\left[\begin{array}{cc}
\left(\sigmasqjjs\right)^{2} & \rhojj\sigmasqjjs\sigmasqjjn\\
\rhojj\sigmasqjjs\sigmasqjjn & \left(\sigmasqjjn\right)^{2}
\end{array}\right].\label{eq:covariance_matrix_sq}
\end{equation}
The parameters of the speech amplitude distribution of each component,
$p\left(\tilde{a}_{n}^{(\tilde{g},\,\breve{g})}|\mathcal{Y}_{n}\right)$,
are obtained using (\ref{eq:omega-nakagami}) and (\ref{eq:m_nakagami})
as 
\begin{align}
\tilde{\Omega}_{n|n}^{\left(\tilde{g},\,\breve{g}\right)} & =\tilde{\mu}_{\mathsf{sq}}^{\left(\tilde{g},\,\breve{g}\right)}\label{eq:omage-calculation}\\
\tilde{m}_{n|n}^{\left(\tilde{g},\,\breve{g}\right)} & =\tilde{\Omega}_{n|n}^{2\left(\tilde{g},\,\breve{g}\right)}/\left(\sigmasqjjs\right)^{2}.\label{eq:m-calculation}
\end{align}
The parameters of the noise amplitude distribution, $p\left(\breve{a}_{n}^{\left(\tilde{g},\,\breve{g}\right)}|\mathcal{Y}_{n}\right)$,
can be estimated from $\breve{\mu}_{\mathsf{sq}}^{\left(\tilde{g},\,\breve{g}\right)}$
and $\left(\sigmasqjjn\right)^{2}$ in the same manner. As a result,
the mean of the amplitudes of speech and noise, $\tilde{\mu}_{n|n}^{\left(\tilde{g},\,\breve{g}\right)}$
and $\breve{\mu}_{n|n}^{\left(\tilde{g},\,\breve{g}\right)}$, can
be calculated using (\ref{eq:mean-nakagami}). Also, the variance
of the speech and noise amplitudes, $\left(\tilde{\sigma}_{n|n}^{\left(\tilde{g},\,\breve{g}\right)}\right)^{2}$
and $\left(\breve{\sigma}_{n|n}^{\left(\tilde{g},\,\breve{g}\right)}\right)^{2}$,
can be calculated using (\ref{eq:variance-nakagami}).

The remaining task is the calculation of the covariance for the speech
and noise amplitude of each Gaussian component, {\small{}$\omega^{\left(\tilde{g},\,\breve{g}\right)}\triangleq\text{\ensuremath{\mathbb{E}}}\left(\tilde{A}_{n}^{\left(\tilde{g},\,\breve{g}\right)},\breve{A}_{n}^{\left(\tilde{g},\,\breve{g}\right)}|\mathcal{Y}_{n}\right)-\text{\ensuremath{\mathbb{E}}}\left(\tilde{A}_{n}^{\left(\tilde{g},\,\breve{g}\right)}|\mathcal{Y}_{n}\right)\text{\ensuremath{\mathbb{E}}}\left(\breve{A}_{n}^{\left(\tilde{g},\,\breve{g}\right)}|\mathcal{Y}_{n}\right)$}.
For two Nakagami-m variables with different parameters $m$, there
is no analytical solution for calculating the correlation coefficient,
$\rho^{\left(\tilde{g},\,\breve{g}\right)}=\frac{\text{\ensuremath{\mathbb{E}}}\left(\widetilde{A}_{n}^{\left(\tilde{g},\,\breve{g}\right)},\breve{A}_{n}^{\left(\tilde{g},\,\breve{g}\right)}|\mathcal{Y}_{n}\right)-\text{\ensuremath{\mathbb{E}}}\left(\widetilde{A}_{n}^{\left(\tilde{g},\,\breve{g}\right)}|\mathcal{Y}_{n}\right)\text{\ensuremath{\mathbb{E}}}\left(\breve{A}_{n}^{\left(\tilde{g},\,\breve{g}\right)}|\mathcal{Y}_{n}\right)}{\sqrt{\text{Var}\left(\widetilde{A}_{n}^{\left(\tilde{g},\,\breve{g}\right)}|\mathcal{Y}_{n}\right)\text{Var}\left(\breve{A}_{n}^{\left(\tilde{g},\,\breve{g}\right)}|\mathcal{Y}_{n}\right)}}$.
However, $\rho^{\left(\tilde{g},\,\breve{g}\right)}$ can be well-approximated
by the correlation coefficient between the squared Nakagami-m variables
\cite{Song2002}, which is given by $\rhojj$ in (\ref{eq:covariance_matrix_sq}).
Thus, we can obtain that $\omega^{\left(\tilde{g},\,\breve{g}\right)}\approx\rho_{\mathsf{sq}}^{\left(\tilde{g},\,\breve{g}\right)}\tilde{\sigma}_{n|n}^{\left(\tilde{g},\,\breve{g}\right)}\breve{\sigma}_{n|n}^{\left(\tilde{g},\,\breve{g}\right)}$
and the covariance matrix, $\boldsymbol{\Sigma}_{n|n}^{\left(\tilde{g},\,\breve{g}\right)}$,
is thereby given by {\small{}$\boldsymbol{\Sigma}_{n|n}^{\left(\tilde{g},\,\breve{g}\right)}=\left[\begin{array}{cc}
\tilde{\sigma}_{n|n}^{2\left(\tilde{g},\,\breve{g}\right)} & \omega^{\left(\tilde{g},\,\breve{g}\right)}\\
\omega^{\left(\tilde{g},\,\breve{g}\right)} & \breve{\sigma}_{n|n}^{2\left(\tilde{g},\,\breve{g}\right)}
\end{array}\right]$}. 

Finally, given the mean and covariance of each Gaussian component,
the posterior estimate of the speech and noise amplitudes required
in (\ref{eq:state-vector-update}) is given by

{\small{}
\begin{equation}
\boldsymbol{\mu}_{n|n}=\underset{\tilde{g},\breve{g}}{\sum}\epsilon_{n|n}^{\left(\tilde{g},\,\breve{g}\right)}\boldsymbol{\mu}_{n|n}^{\left(\tilde{g},\,\breve{g}\right)}=\underset{\tilde{g},\breve{g}}{\sum}\epsilon_{n|n}^{\left(\tilde{g},\,\breve{g}\right)}\left[\tilde{\mu}_{n|n}^{\left(\tilde{g},\,\breve{g}\right)},\breve{\mu}_{n|n}^{\left(\tilde{g},\,\breve{g}\right)}\right]^{\mathsf{T}},\label{eq:mean-amplitude}
\end{equation}
}and the covariance matrix in required in (\ref{eq:covariance-matrix-update})
is given by

{\small{}
\begin{equation}
\boldsymbol{\Sigma}_{n|n}=\underset{\tilde{g},\breve{g}}{\sum}\epsilon_{n|n}^{\left(\tilde{g},\,\breve{g}\right)}\left(\boldsymbol{\Sigma}_{n|n}^{\left(\tilde{g},\,\breve{g}\right)}+\boldsymbol{\mu}_{n|n}^{\left(\tilde{g},\,\breve{g}\right)}\left(\boldsymbol{\mu}_{n|n}^{\left(\tilde{g},\,\breve{g}\right)}\right)^{\mathsf{T}}\right)-\boldsymbol{\mu}_{n|n}\boldsymbol{\mu}_{n|n}^{\mathsf{T}}.\label{eq:variance-amplitude}
\end{equation}
}{\small \par}

In this section, the entire process of calculating the posterior estimate
of both speech and noise from their prior estimate. has been described.
First, the parameters of the Nakagami-m distribution are calculated
by fitting to the prior estimate of speech and noise using (\ref{eq:omega_naka})
and (\ref{eq:m-naka}) and get the parameters of the corresponding
Rician distribution from them using (\ref{eq:alpha-Rician}) and (\ref{eq:epslion-Rician}).
Thus, the mean and covariance of each Gaussian component are obtained
from (\ref{eq:mean-real}) to (\ref{eq:variance-real-imag}) and the
posterior distribution of the Gaussring components is obtained as
the pairwise product of the components of speech and noise. Second,
the parameters of the amplitude distribution for each component of
the posterior distribution are calculated using (\ref{eq:omage-calculation})
and (\ref{eq:m-calculation}). Given these parameters, the mean vector
and the covariance matrix of the speech and noise amplitudes, namely
$\boldsymbol{\mu}_{n|n}^{\left(\tilde{g},\,\breve{g}\right)}$ and
$\boldsymbol{\Sigma}_{n|n}^{\left(\tilde{g},\,\breve{g}\right)}$,
can be calculated for each Gaussian component. Finally, the overall
mean vector, $\boldsymbol{\mu}_{n|n}$, and the covariance matrix,
$\boldsymbol{\Sigma}_{n|n}$, of the posterior estimate are obtained
using (\ref{eq:mean-amplitude}) and (\ref{eq:variance-amplitude}),
respectively. 

\section{Implementation and evaluation\label{sec:Implementation-and-evaluation}}

In this section, the proposed modulation-domain Kalman filter based
MMSE estimator using the update in Sec.~\ref{subsec:Generalized-Gamma-Speech}
is denoted as MDKM and that using the Gaussring-based update in Sec.~\ref{sec:Enhancement-with-Gaussring-Priors}
is denoted as MDKR. The performance of the MDKM and MDKR enhancers
are compared with that of a baseline logMMSE enhancer \cite{Ephraim1985,Brookes1998-2012a},
of a deep neural network (DNN) based enhancer \cite{Wang2014} and
of the colored-noise version of the modulation Kalman filter (MDKFC)
enhancer from \cite{So2011}. The evaluation metrics comprise segSNR
\cite{Hu2006a}, PESQ \cite{Rix2006}, the short-time objective intelligibility
(STOI) measure \cite{Taal2011} and the phone error rate (PER) from
an automatic speech recognition (ASR) system. For the DNN based enhancer,
a DNN was trained to estimate the ideal ratio mask (IRM) \cite{Wang2014}
and it had three 1024-dimensional hidden layers with rectified linear
units (ReLU) \cite{Zeiler2013}. Sigmoid activation functions were
applied in the output layer since the targets are in the range $[0,1]$.
The average mean square error (MSE) between the predicted and true
IRM was used as the cost function. We used an adaptive gradient descent
algorithm \cite{Duchi2011} with a momentum of 0.5. For training the
DNN, 2000 utterances were randomly selected from TIMIT training set
as in \cite{Wang2014} and they were corrupted by babble, factory,
car and destroyer engine noise from the RSG-10 database \cite{Steeneken1988}
at $-10$, $-5$, $0$, $5$, $10$ and $15$\,dB global SNR. The
input features set was same as that in \cite{Wang2014}, which included
amplitude modulation spectrogram, relative spectral transformed perceptual
linear prediction coefficients (RASTA-PLP), mel-frequency cepstral
coefficients (MFCC) and 64-channel Gammatone filterbank power spectra.

\begin{figure}
\begin{centering}
\includegraphics[scale=0.36]{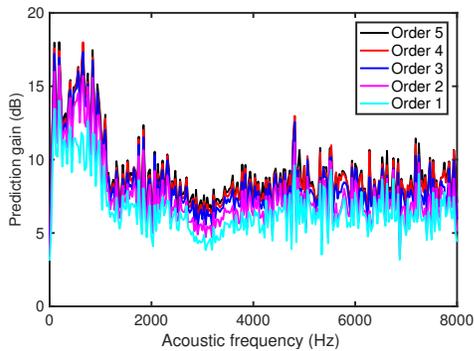}
\par\end{centering}
\caption{Prediction gain for speech modulation-domain LPC model of different
orders.\label{fig:Prediction-gain-speech}}
\end{figure}

\begin{figure}
\begin{centering}
\includegraphics[scale=0.36]{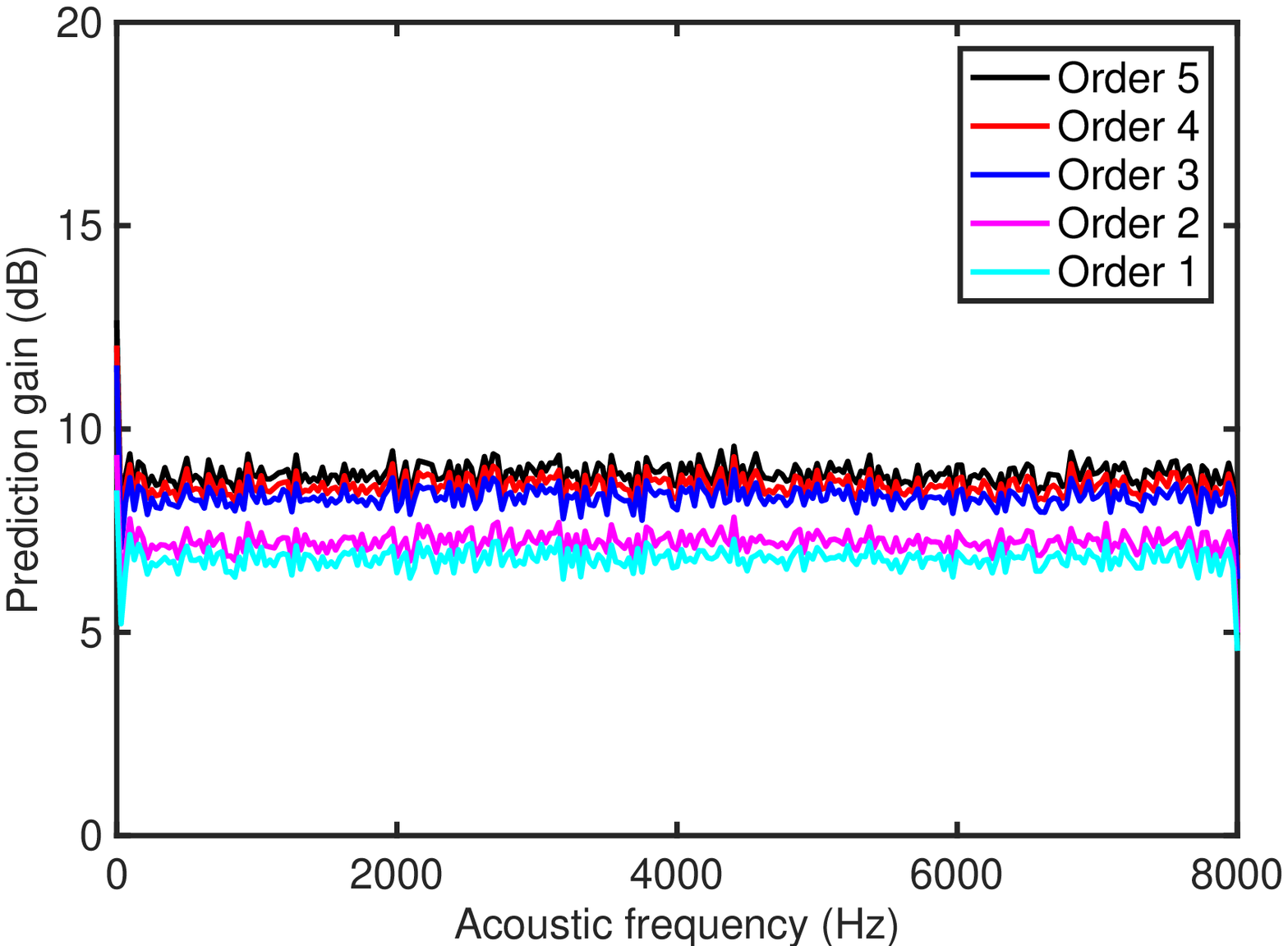}\\
\includegraphics[scale=0.36]{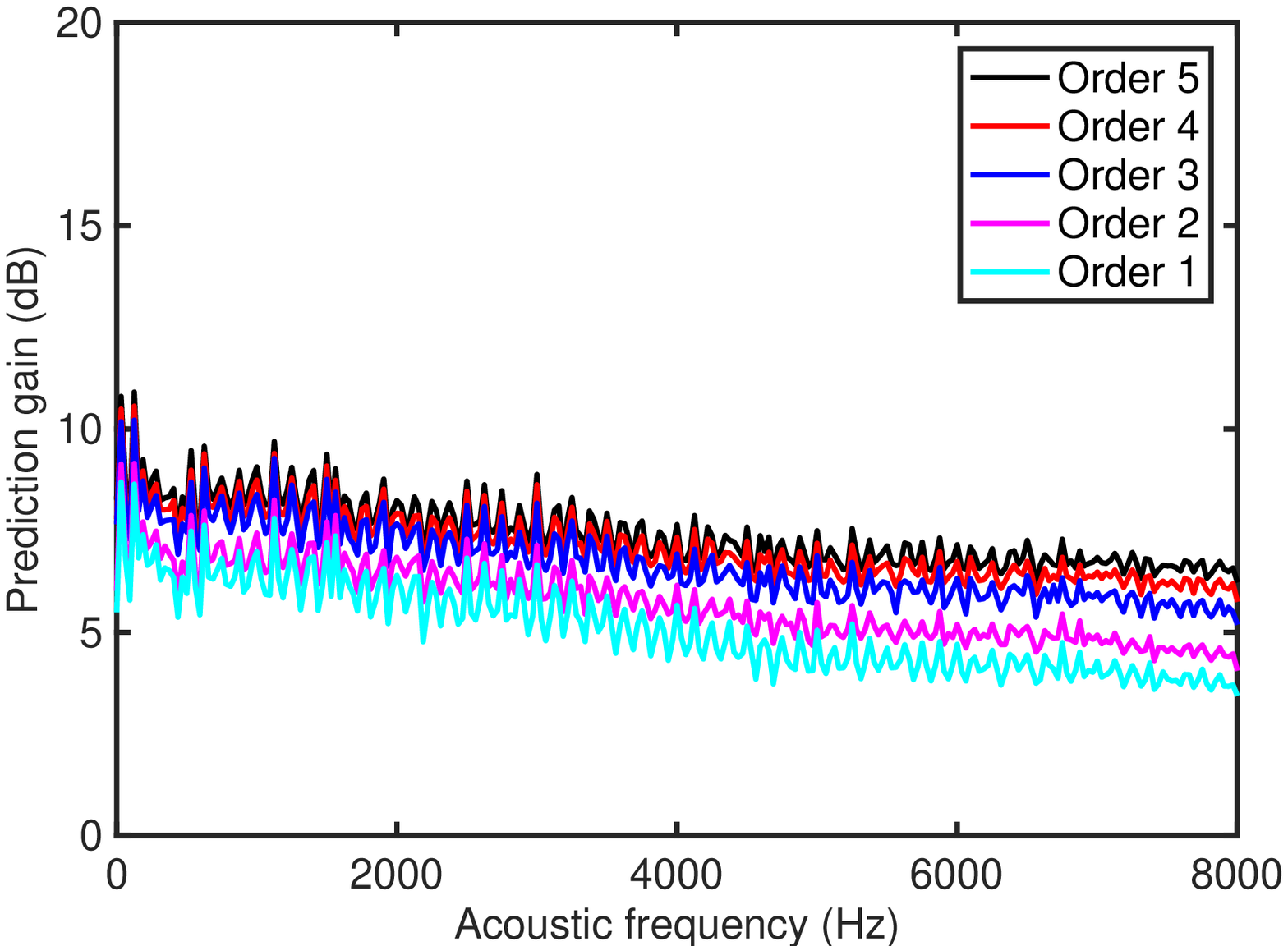}\\
\includegraphics[scale=0.36]{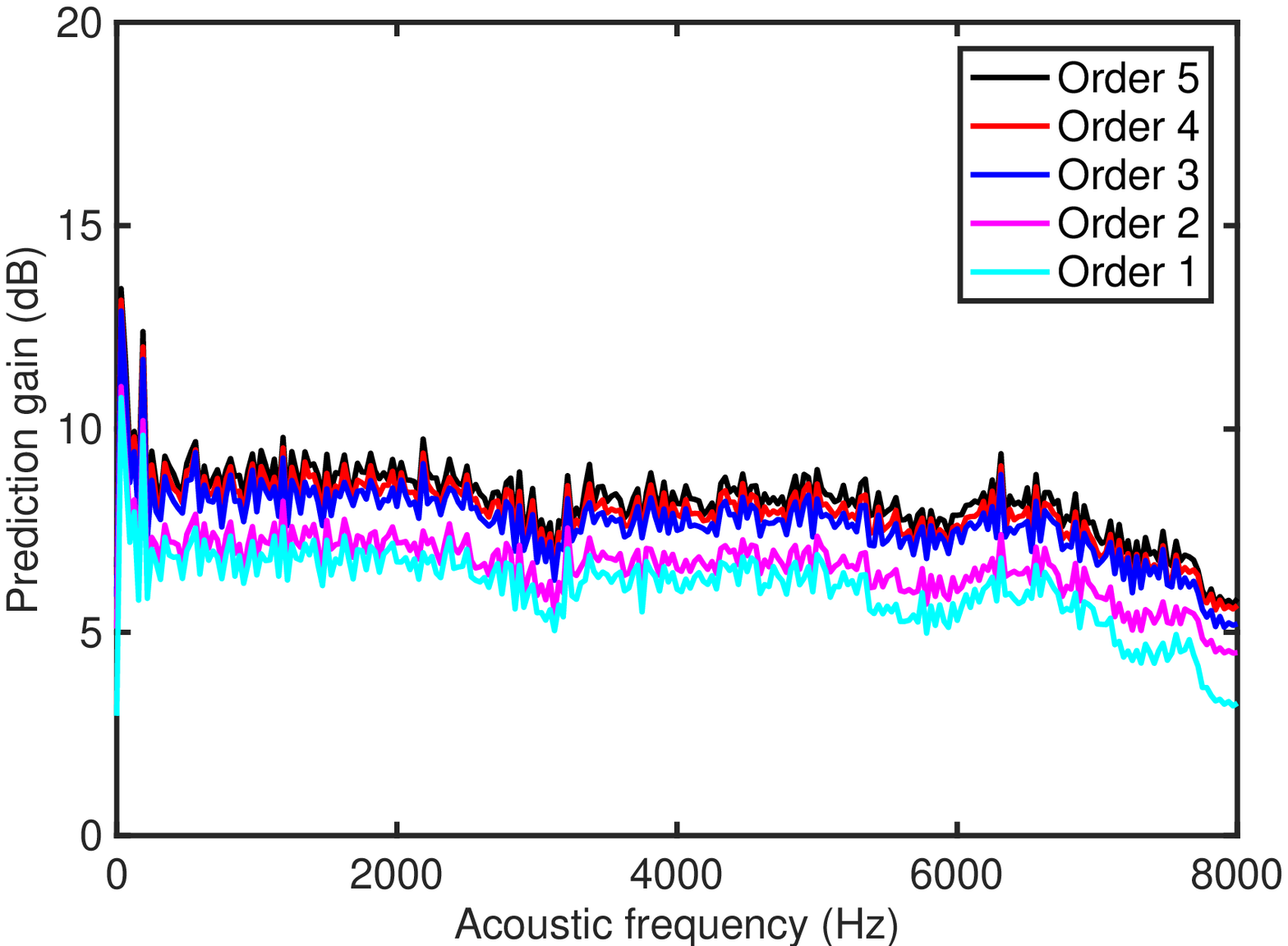}
\par\end{centering}
\caption{Prediction gain for modulation-domain LPC models of different orders
of white noise (top), car noise (middle) and street noise (bottom).\label{fig:Prediction-gain-for-noise}}
\end{figure}

\begin{table}
\caption{Parameter settings in the experiments.\label{tab:Parameter-settings}}
\centering{}{\small{}}%
\begin{tabular}{|cc|}
\hline 
{\small{}Parameter} & {\small{}Settings}\tabularnewline
\hline 
\hline 
{\small{}Sampling frequency} & \selectlanguage{british}%
{\small{}16\,kHz}\selectlanguage{english}%
\tabularnewline
{\small{}Speech/Noise Acoustic frame length} & {\small{}$32$}\foreignlanguage{british}{{\small{}\,}}{\small{}ms}\tabularnewline
{\small{}Speech/Noise Acoustic frame increment} & {\small{}$8$}\foreignlanguage{british}{{\small{}\,}}{\small{}ms}\tabularnewline
{\small{}Speech modulation frame length} & {\small{}$64$}\foreignlanguage{british}{{\small{}\,}}{\small{}ms}\tabularnewline
{\small{}Speech modulation frame increment} & {\small{}$8$}\foreignlanguage{british}{{\small{}\,}}{\small{}ms}\tabularnewline
{\small{}Noise modulation frame length} & {\small{}$64$}\foreignlanguage{british}{{\small{}\,}}{\small{}ms}\tabularnewline
{\small{}Noise modulation frame increment} & {\small{}$16$}\foreignlanguage{british}{{\small{}\,}}{\small{}ms}\tabularnewline
{\small{}Analysis-synthesis window} & {\small{}Hamming window}\tabularnewline
{\small{}Speech LPC model order $p$} & {\small{}$3$}\tabularnewline
{\small{}noise LPC model order $q$} & {\small{}$4$}\tabularnewline
\hline 
\end{tabular}{\small \par}
\end{table}

The evaluations used the core test set from the TIMIT database \cite{Garofolo1988}
as the test set, which contains 16 male and 8 female speakers each
reading 8 sentences for a total of 192 sentences all with distinct
texts. In order to optimize the parameters of the algorithms other
than the LPC orders, a development set was used that comprised of
200 speech sentences randomly selected from the development set of
the TIMIT database. A summary of the parameter settings is given in
Table~\ref{tab:Parameter-settings}. The speech was corrupted by
F16 noise from the RSG-10 database \cite{Steeneken1988} and street
noise from the ITU-T test signals database \cite{ITU_T_P501}. The
sampling rate of the speech signals was 16\,kHZ and noise signals
were downsampled to 16\,kHz. The speech LPC coefficients for the
MDKM, MDKR and MDKFC algorithms were estimated from each modulation
frame of the logMMSE-enhanced speech. In order to estimate the noise
LPC models for the MDKR and MDKFC algorithms, we followed the procedure
described in \cite{So2011} in which the estimated modulation magnitude
spectrum of the noise was recursively averaged during intervals that
were classified as noise-only. The noise LPC coefficients were then
found from the autocorrelation coefficients of the modulation magnitude
spectrum of the noise. The prediction residual signal of speech and
noise, which were denoted as $\tilde{\eta}^{2}$ and $\breve{\eta}^{2}$
in $\mathbf{Q}_{n}$ in (\ref{eq:covariance_prediction}), were calculated
as the power of the prediction errors for each modulation frame. To
investigate the effect of the order on the speech modulation-domain
LPC model, we calculated the prediction gain for a range of LPC orders.
The prediction gain, $\Xi_{\mathsf{p}}$, is defined as 

\begin{equation}
\Xi_{\mathsf{p}}\triangleq\frac{\mathbb{E}\left(|S_{n,k}|^{2}\right)}{\mathbb{E}\left(\left(|S_{n,k}|-|\widehat{S}_{n,k}|\right)^{2}\right)}\label{eq:prediction-gain}
\end{equation}
where $|\widehat{S}_{n,k}|$ represents the estimated speech amplitude.
The expectation in (\ref{eq:prediction-gain}) was taken over all
acoustic frames for each frequency bin. In Fig.~\ref{fig:Prediction-gain-speech},
we show the prediction gain of clean speech which was formed using
$100$ speech sentences from the development set. From Fig.~\ref{fig:Prediction-gain-speech},
it can be seen that, when the order, $p$, of the modulation-domain
LPC model is $\geq$~$2$ , the prediction gain exceeds 10\,dB at
most acoustic frequencies. For the acoustic frequencies accounting
for most of the speech power ($500-1000$\,Hz), the prediction gain
exceeds $15$\,dB. In the evaluation experiments, a modulation-domain
LPC model of order 3 was used when a speech LPC model was required.
Similarly, Fig.~\ref{fig:Prediction-gain-for-noise} shows the prediction
gain of the noise LPC model for different orders, $q$, for white
noise, car noise and street noise. The plots show that the LPC models
with of order $\geq$ 3 are able to model the noises in the modulation
domain. The prediction gains of white noise are about $10$\,dB over
acoustic frequencies, which are fairly stable because of the stationary
power distribution of white noise (the sudden drop of prediction gain
at very low and very high frequencies results from the framing and
windowing in the time domain). It worth noting that the predictability
of the spectral amplitudes of the white noise results from the amplitude
correlation that is introduced by the overlapped windows in the STFT.
For car noise, because nearly all of acoustic spectral power is at
low acoustic frequencies, the temporal acoustic sequences within these
frequency bins are easier to predict from the previous acoustic frames,
therefore the prediction gains are clearly higher at low frequencies
than those at high frequencies, which are about $12$\,dB. For the
street noise, the gains are similar to those of the white noise and
car noise. At low frequencies (10 to 200 Hz) the prediction gains
are higher (about $14$\,dB) than those of higher frequencies. In
the experiments, a modulation-domain LPC model of order $4$ was used
when a noise LPC model was required. 

\begin{figure}
\noindent \begin{centering}
\includegraphics[scale=0.4]{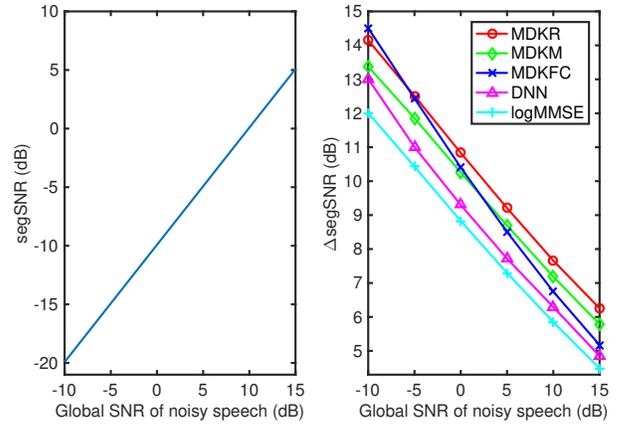}
\par\end{centering}
\caption{Left: Average segmental SNR plotted against the global SNR of the
input speech corrupted by additive F16 noise. Right: Average segmental
SNR improvement after processing by four algorithms plotted against
the global SNR of the input speech corrupted by additive F16 noise.
The algorithm acronyms are defined in the text. \label{fig:segSNR-Kring-car}}
\end{figure}

\begin{figure}
\noindent \begin{centering}
\includegraphics[scale=0.4]{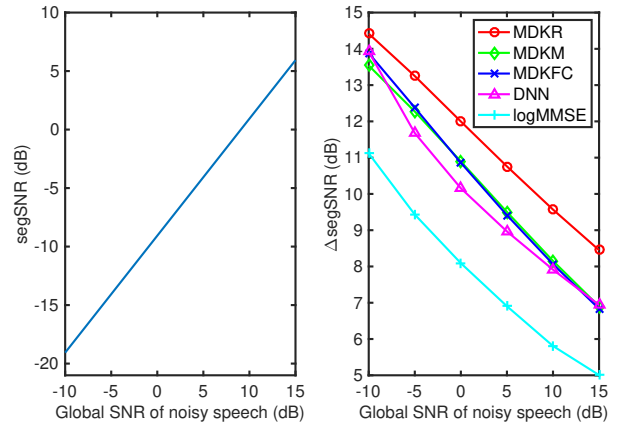}
\par\end{centering}
\caption{Left: Average segmental SNR plotted against the global SNR of the
input speech corrupted by additive street noise. Right: Average segmental
SNR improvement after processing by four algorithms plotted against
the global SNR of the input speech corrupted by additive street noise.
\label{fig:segSNR-Kring-street}}
\end{figure}

\begin{figure}
\noindent \begin{centering}
\includegraphics[scale=0.4]{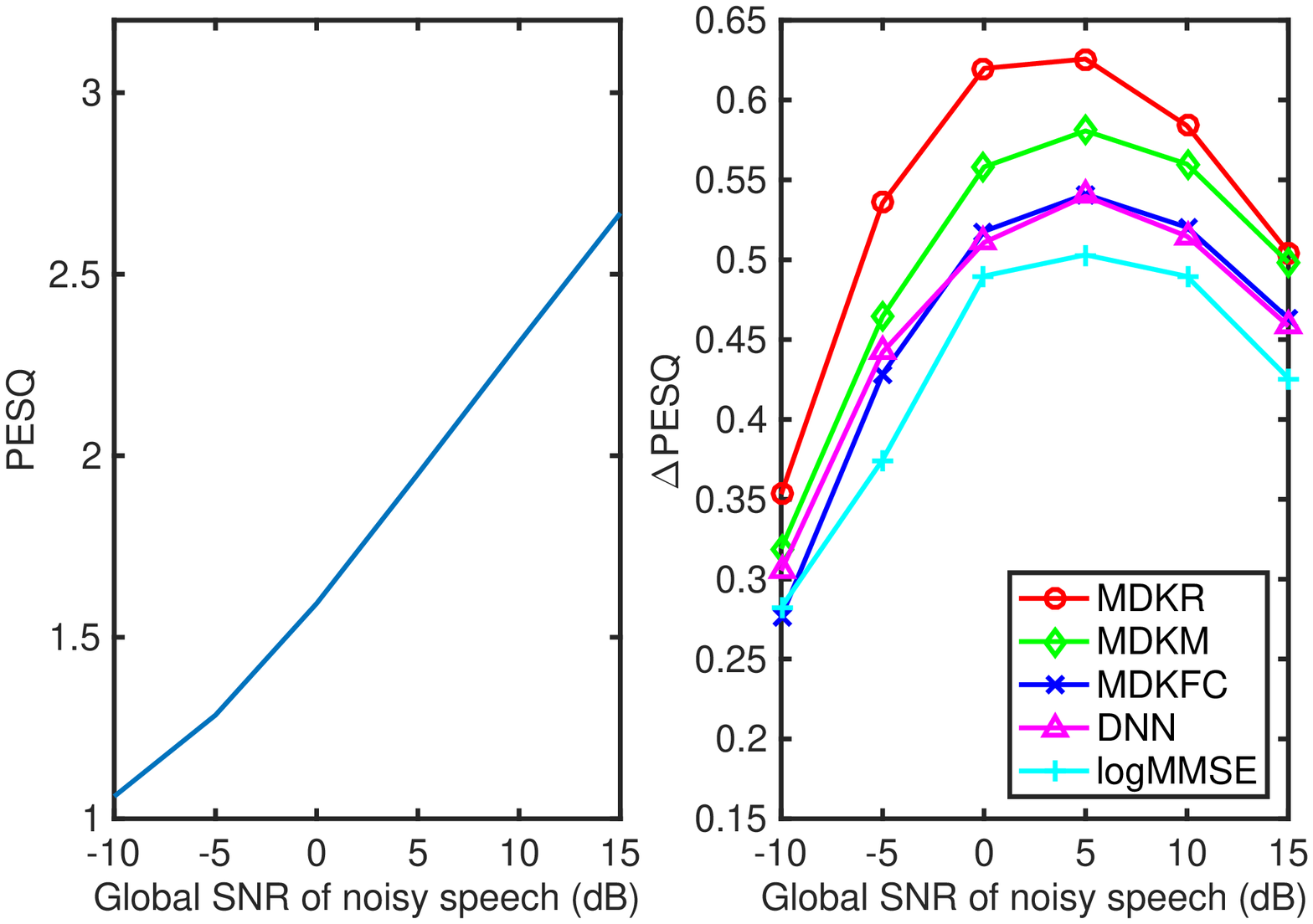}
\par\end{centering}
\caption{Left: Average PESQ plotted against the global SNR of the input speech
corrupted by additive F16 noise. Right: Average PESQ of enhanced speech
after processing by four algorithms plotted against the global SNR
of the input speech corrupted by additive F16 noise. \label{fig:PESQ-Kring-car}}
\end{figure}

\begin{figure}
\noindent \begin{centering}
\includegraphics[scale=0.4]{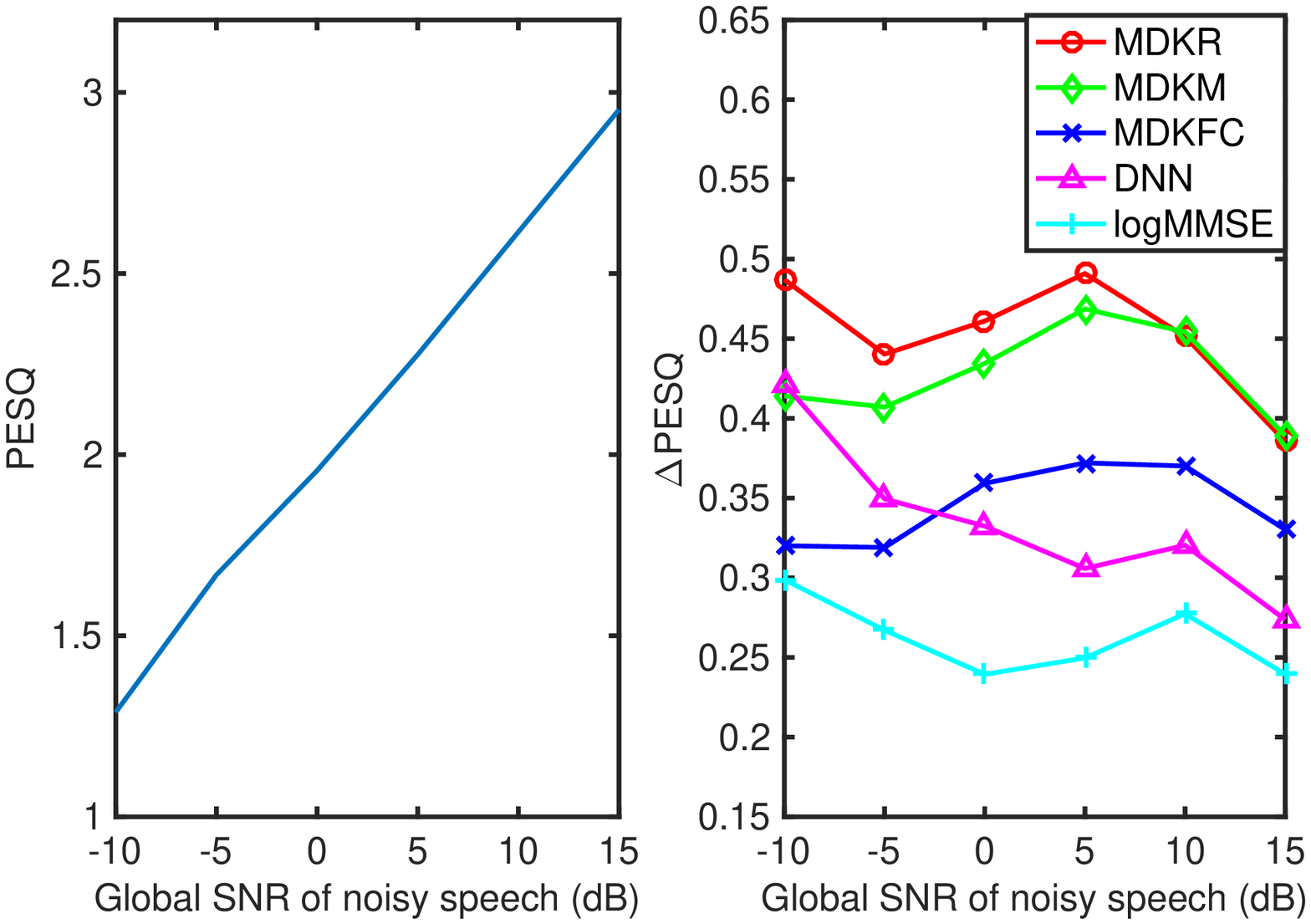}
\par\end{centering}
\caption{Left: Average PESQ plotted against the global SNR of the input speech
corrupted by additive street noise. Right: Average PESQ of enhanced
speech after processing by four algorithms plotted against the global
SNR of the input speech corrupted by additive street noise.\label{fig:PESQ-Kring-street}}
\end{figure}

\begin{figure}
\noindent \begin{centering}
\includegraphics[scale=0.4]{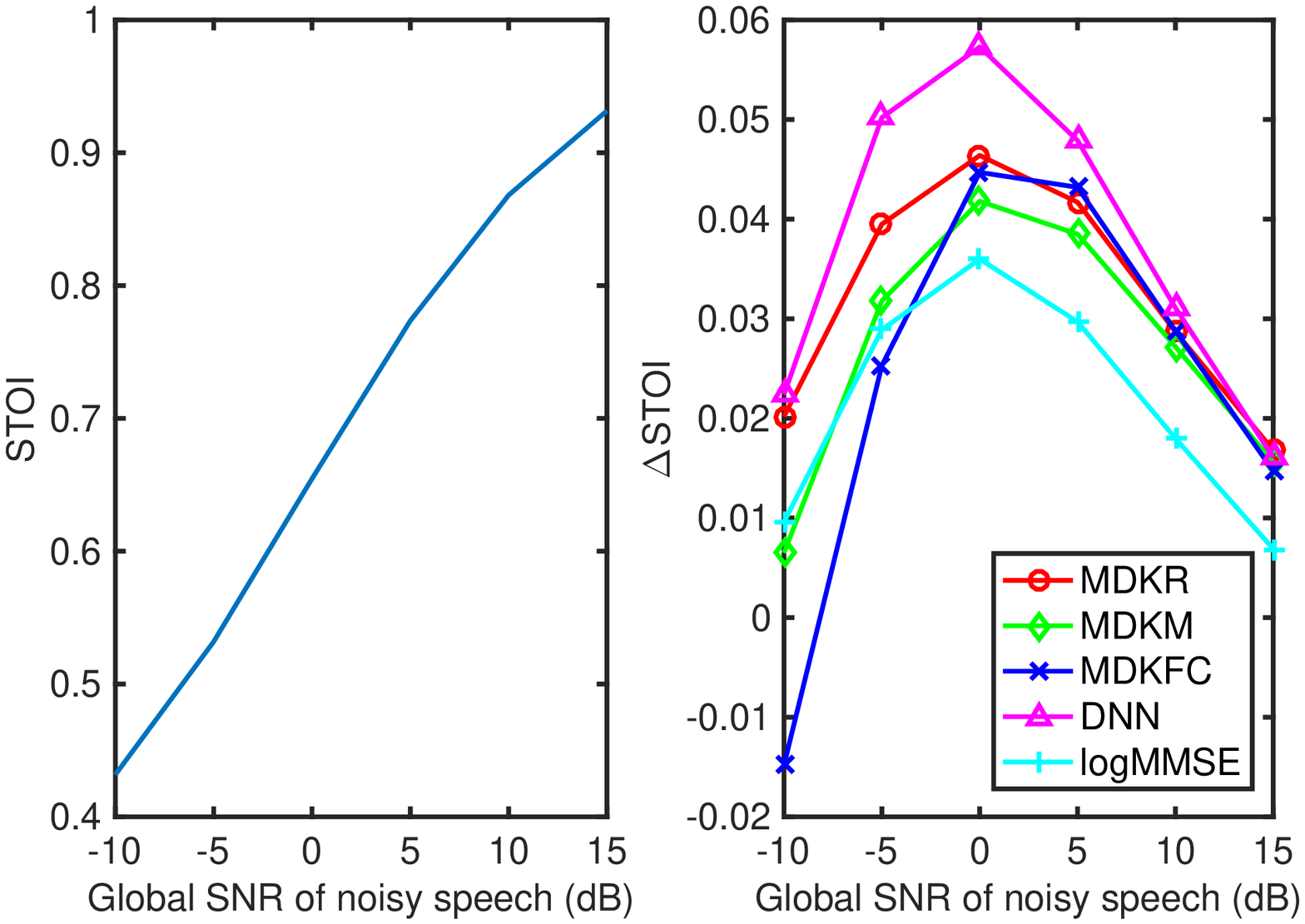}
\par\end{centering}
\caption{Left: Average STOI plotted against the global SNR of the input speech
corrupted by additive F16 noise. Right: Average STOI of enhanced speech
after processing by four algorithms plotted against the global SNR
of the input speech corrupted by additive F16 noise. \label{fig:STOI-Kring-car}}
\end{figure}

\begin{figure}
\noindent \begin{centering}
\includegraphics[scale=0.4]{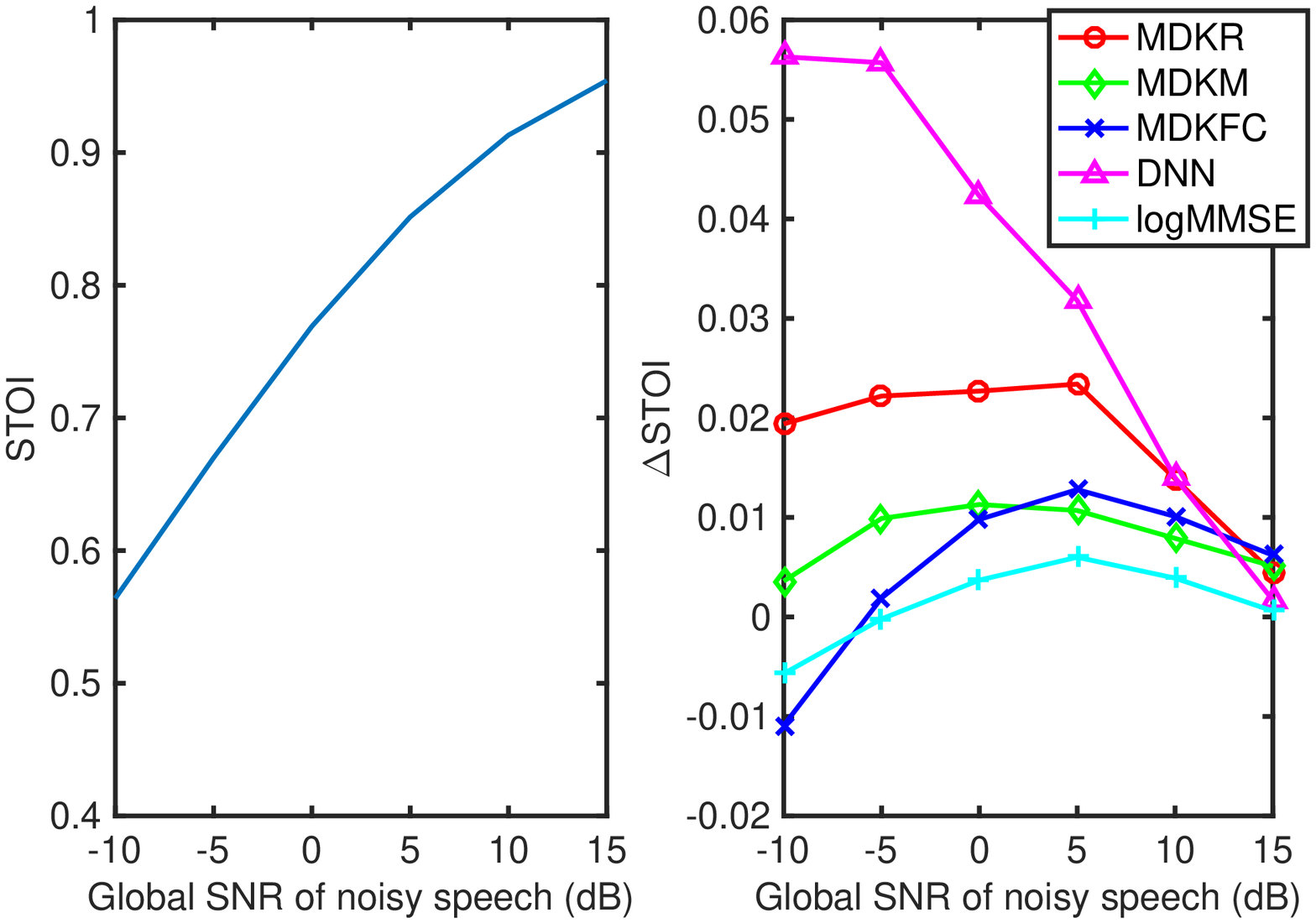}
\par\end{centering}
\caption{Left: Average STOI plotted against the global SNR of the input speech
corrupted by additive street noise. Right: Average STOI of enhanced
speech after processing by four algorithms plotted against the global
SNR of the input speech corrupted by additive street noise.\label{fig:STOI-Kring-street}}
\end{figure}

\begin{figure}
\noindent \begin{centering}
\includegraphics[scale=0.4]{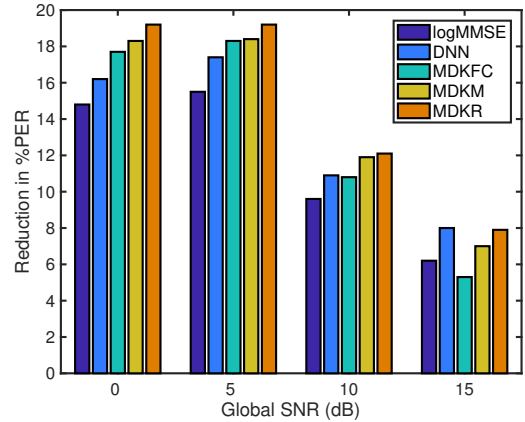}
\par\end{centering}
\caption{Phone Error Rate (PER) reduction plotted against the global SNR of
the input speech corrupted by additive F16 noise. The PERs of the
noisy speech at \{$0,5,10,15$\}\,dB SNR were \{$86.7,71.7,52.2,38.3$\}$\%$
respectively. \label{fig:ASR_f16}}
\end{figure}

\begin{figure}
\noindent \begin{centering}
\includegraphics[scale=0.4]{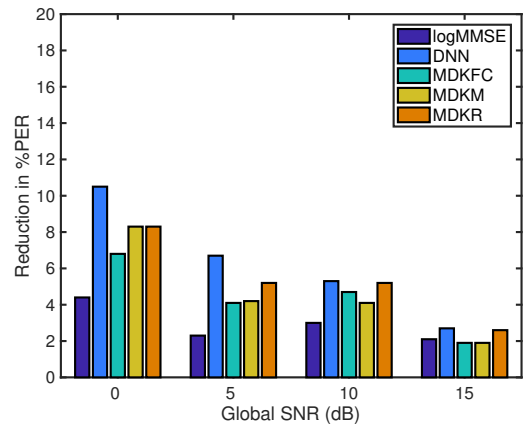}
\par\end{centering}
\caption{Phone Error Rate (PER) reduction plotted against the global SNR of
the input speech corrupted by additive street noise. The PERs of the
noisy speech at \{$0,5,10,15$\}\,dB SNR were \{$64.3,47.9,36.9,28.1$\}$\%$
respectively. \label{fig:ASR_street}}
\end{figure}

The speech signals were corrupted with additive F16 noise from the
RSG-10 database \cite{Steeneken1988} and street noise \cite{ITU_T_P501}
at $-10$,$-5$, $0$, $5$, $10$ and $15$\,dB global SNR. All
the measured values shown are averages over all the sentences in the
TIMIT core test set. Figures \ref{fig:segSNR-Kring-car} and \ref{fig:segSNR-Kring-street}
show the average segSNR of the noisy speech and the average segSNR
improvement given by each algorithm over the noisy speech at each
SNR for F16 noise and street noise, respectively. It can be seen that,
for F16 noise, the MDKFC algorithm performs better than the MDKR,
MDKM and DNN enhancers at -10\,dB SNRs while at high SNRs, the MDKFR
enhancer outperforms MDKFC by about 1\,dB and MDKM algorithms by
about 0.5\,dB. At -10\,dB, the DNN enhancer performs similarly to
the MDKM enhancer and at other SNRs it performs worse than the MDKM
enhancer by about 1\,dB. For street noise, the MDKFR enhancer gives
an improvement of by 2 to 3\,dB over the MDKM and MDKFC enhancers
over the entire range of SNRs. The DNN enhancer performs slight worse
than the MDKM and MDKFC enhancers and it gives about 2.5\,dB improvement
over the logMMSE enhancer.

Figures \ref{fig:PESQ-Kring-car} and \ref{fig:PESQ-Kring-street}
give the corresponding average PESQ of the noisy speech and the average
PESQ performance improvement over noisy speech at each SNR. It shows
that for F16 noise, at -10\,dB and 15\,dB SNRs, the MDKR, MDKM give
similar performance and at other SNRs, the MDKR enhancer gives an
improvement of about 0.1 over the MDKM and about 0.2 over the logMMSE
enhancer. The MDKFC enhancer performs slightly worse that the MDKM
enhancer and outperforms the logMMSE enhancer by about 0.05. The DNN
enhancer gives a similar performance as the MDKFC enhancer. For street
noise, the MDKR enhancer gives an improvement of around 0.1 over the
MDKM enhancer at -10\,dB SNR and at high SNRs ($>$10\,dB), they
give similar performance. The DNN enhancer gives similar performance
as the MDKM enhancer at -10\,dB. At high SNRs, the performance of
the DNN enhancer is worse than the MDKM and MDKFC enhancer by around
0.15 and 0.05, respectively. 

\begin{figure*}
\begin{centering}
\includegraphics[scale=0.33]{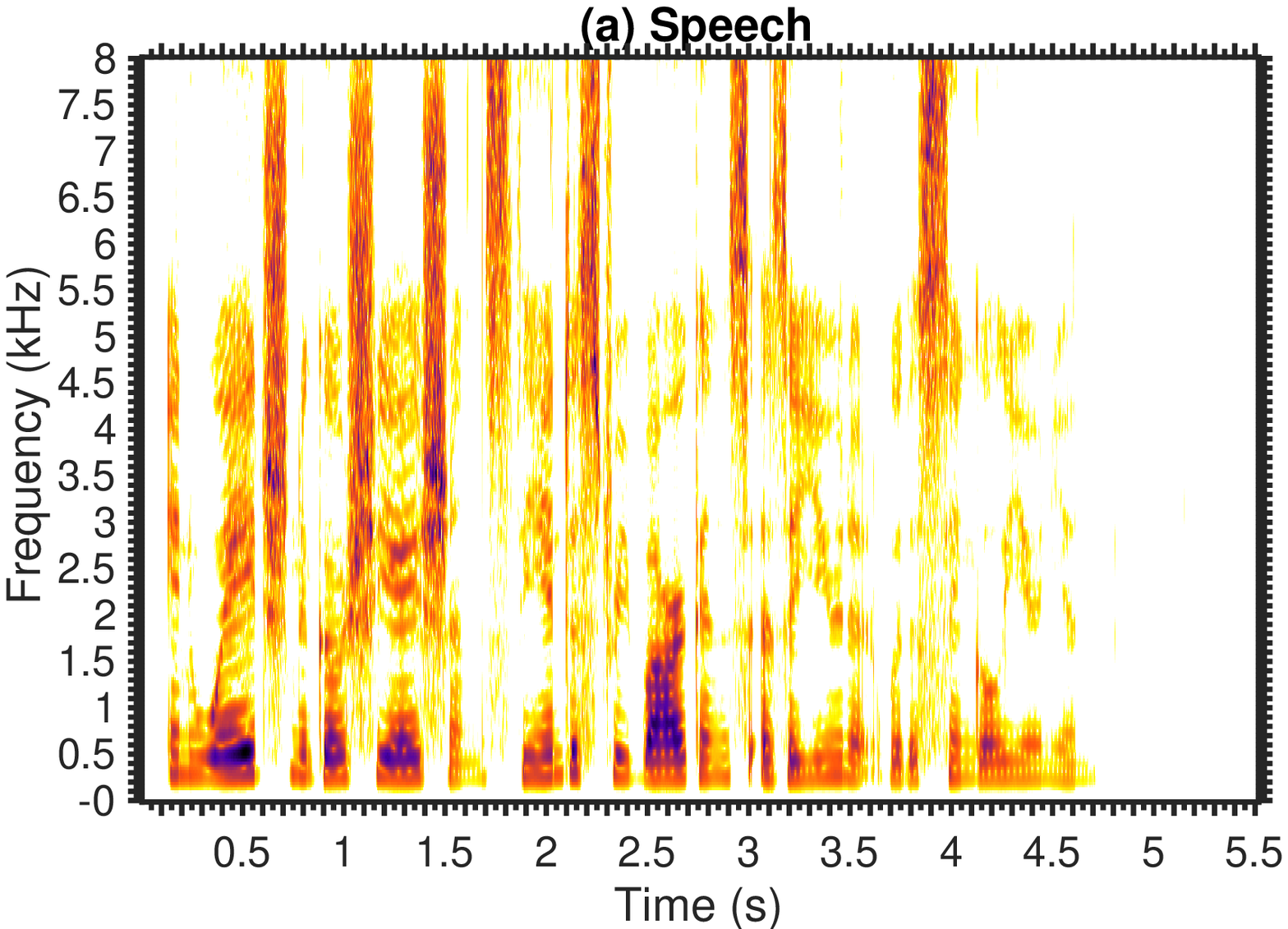}\includegraphics[scale=0.33]{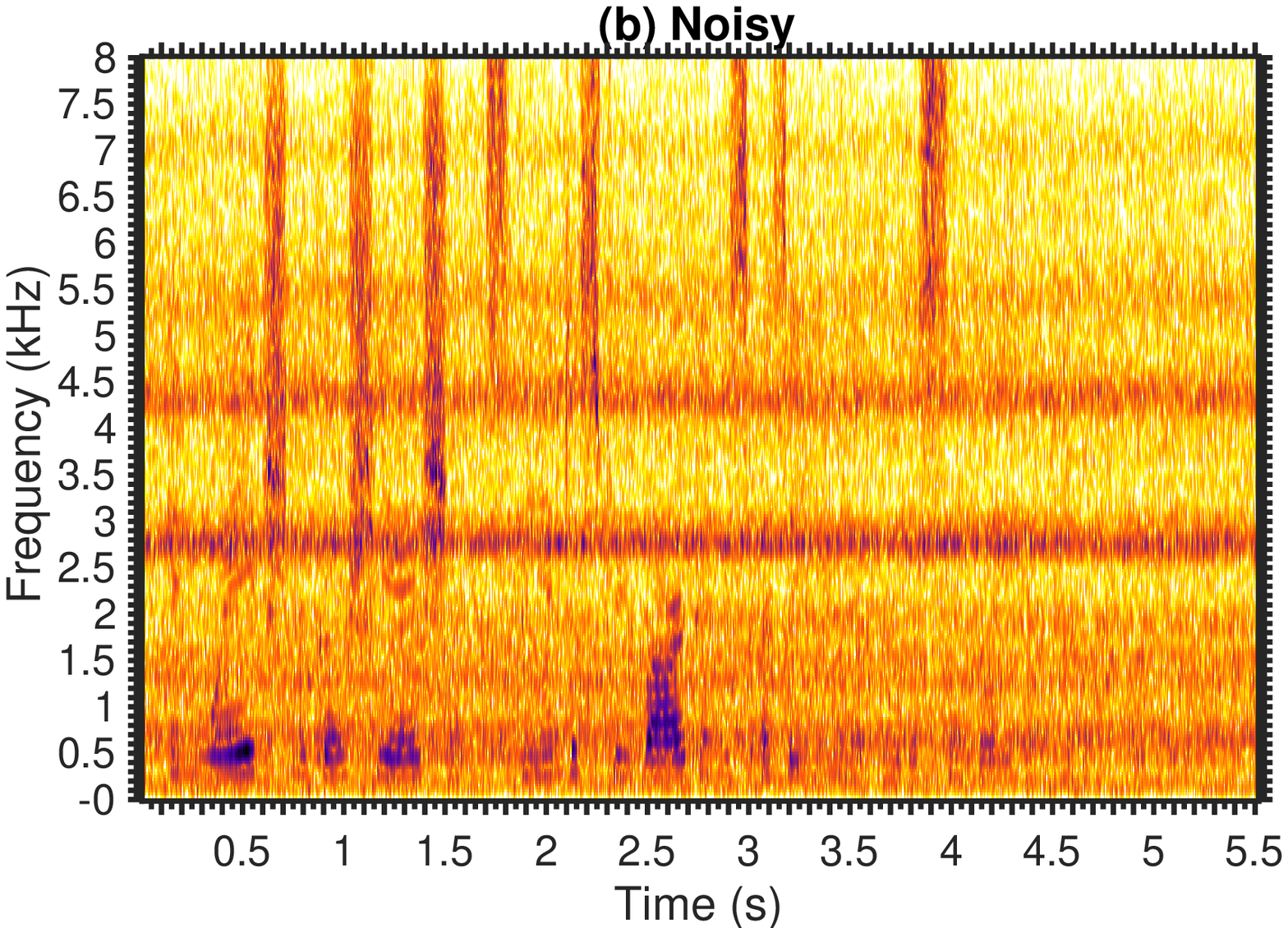}\includegraphics[scale=0.33]{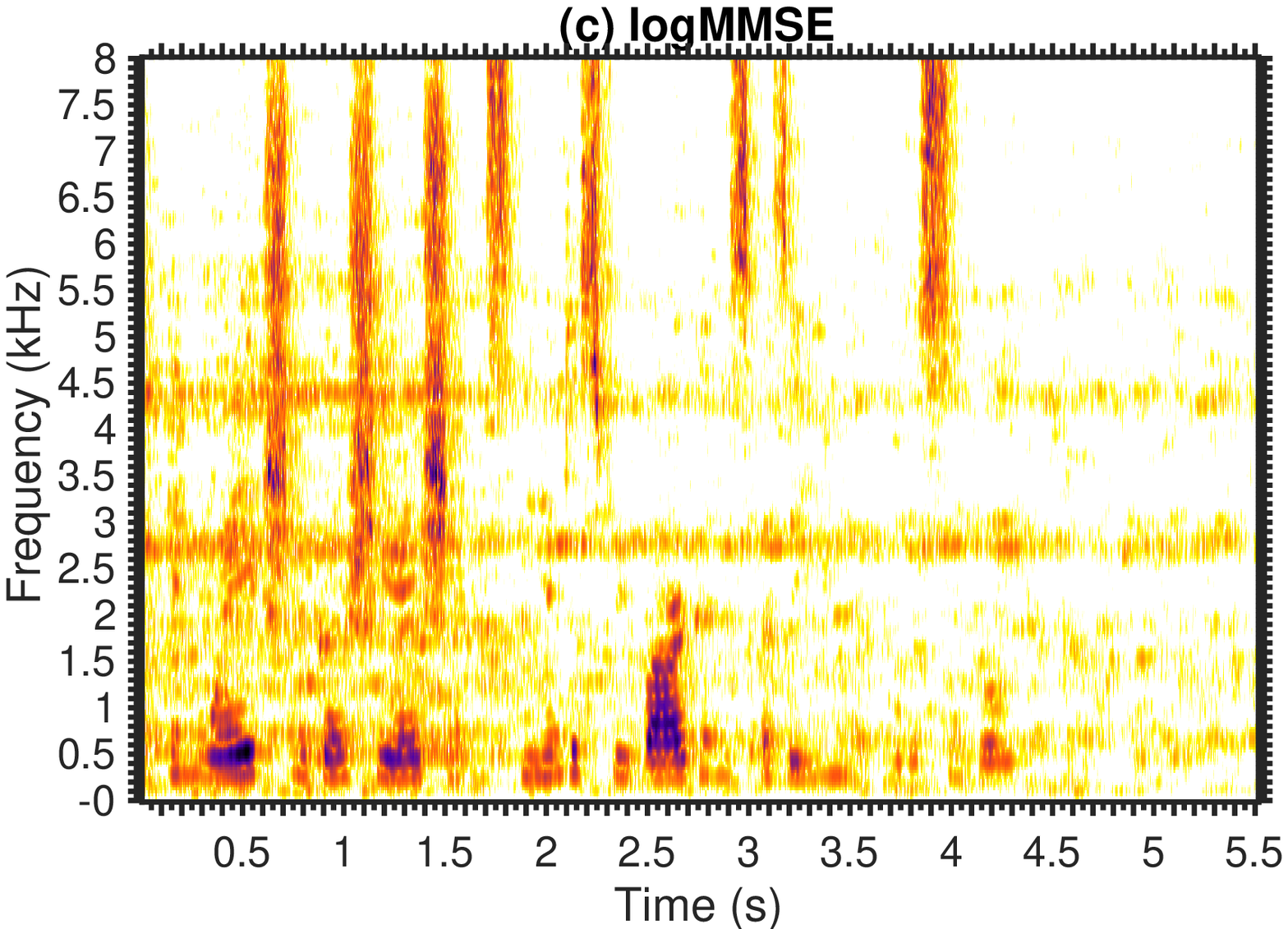}
\par\end{centering}
\begin{centering}
\includegraphics[scale=0.33]{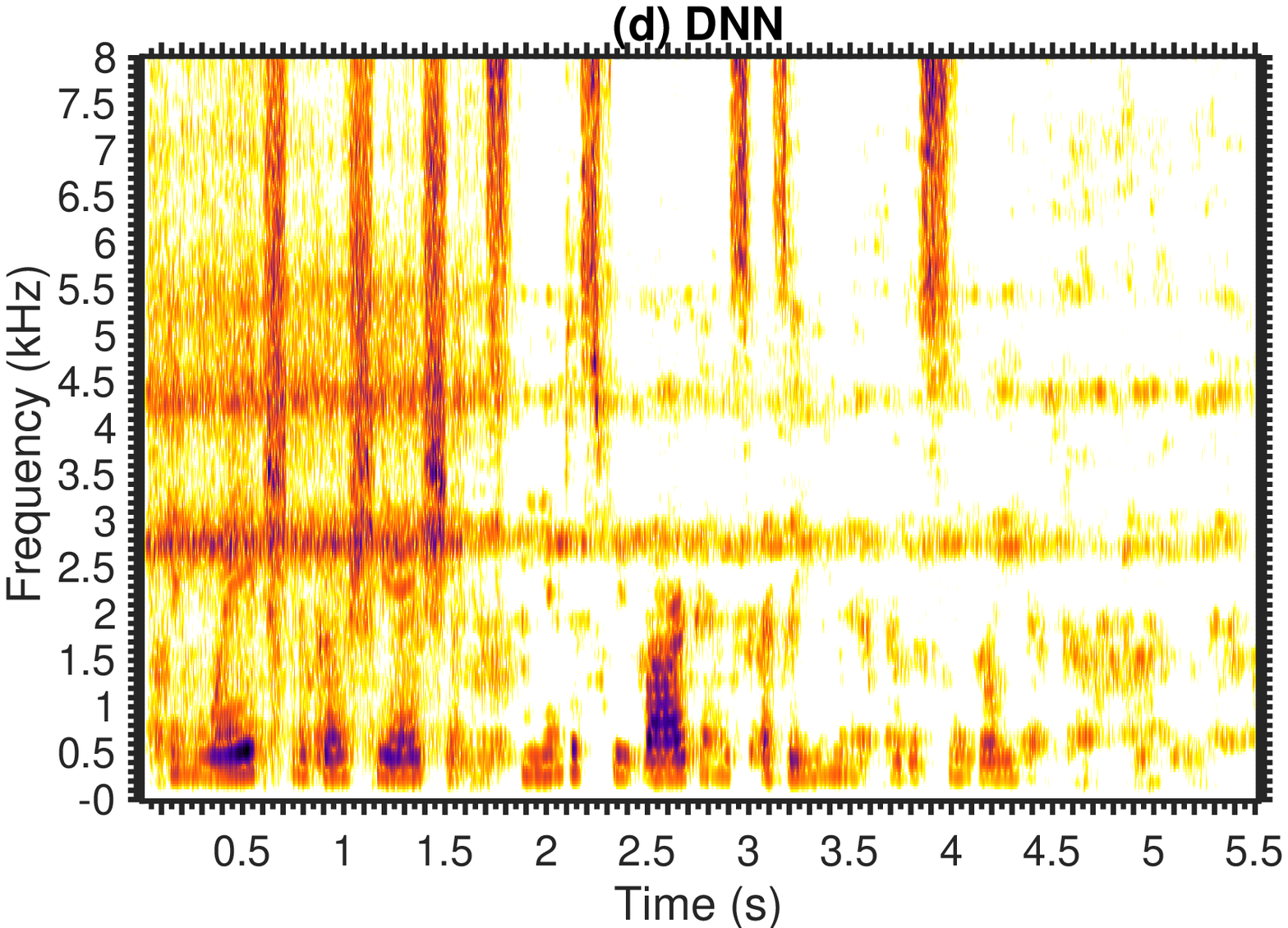}\includegraphics[scale=0.33]{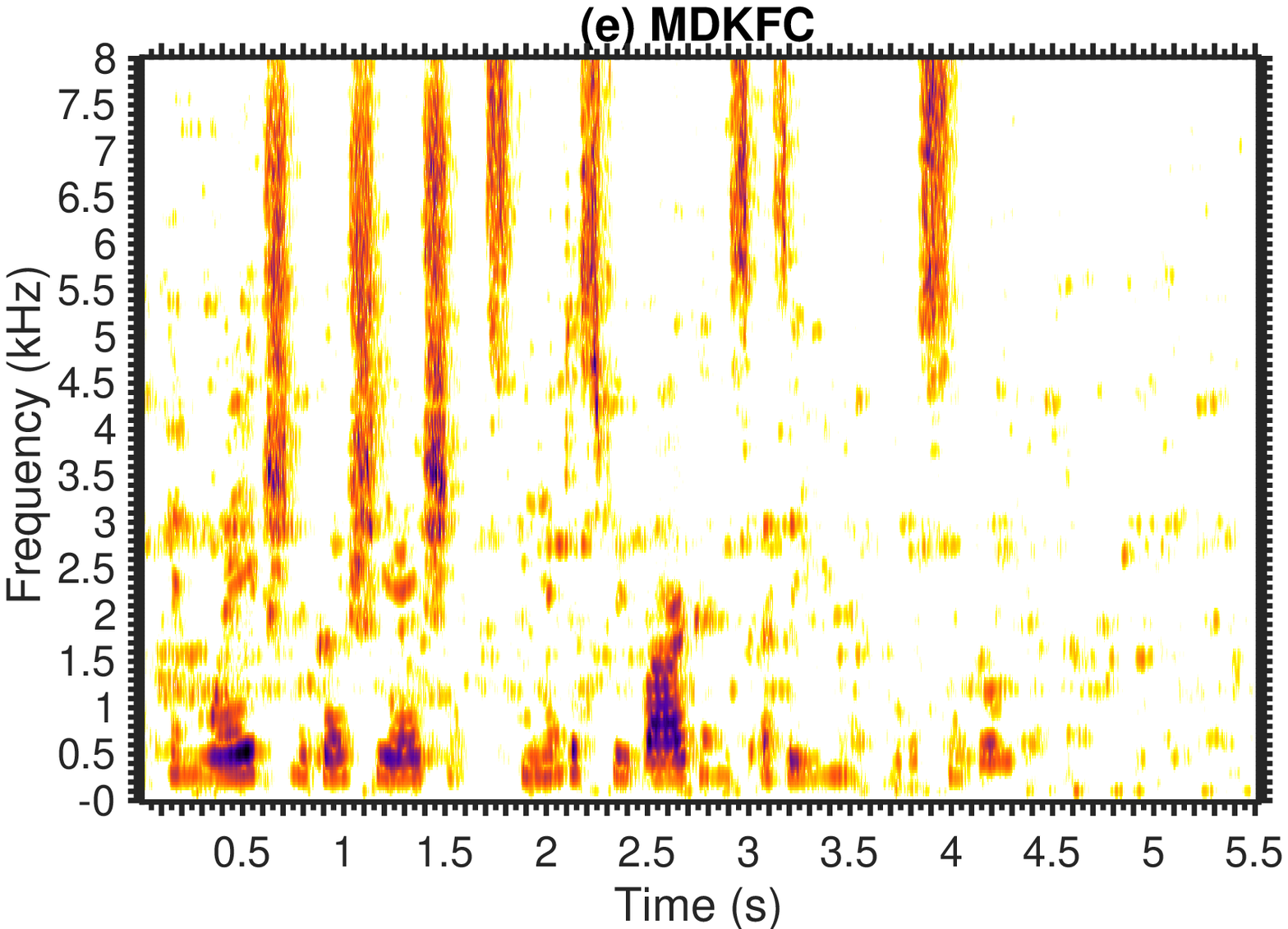}\includegraphics[scale=0.33]{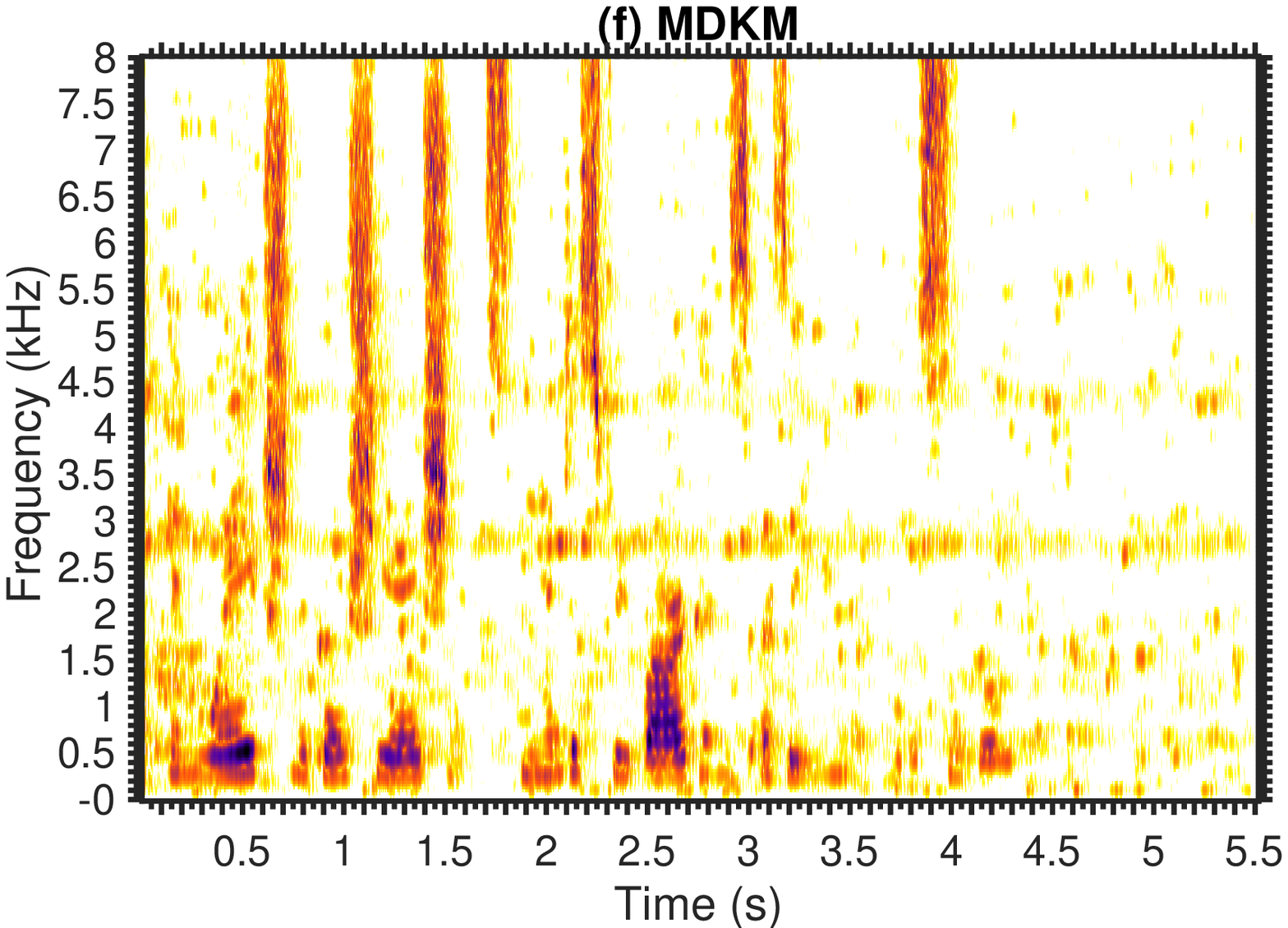}
\par\end{centering}
\begin{centering}
~~~~~~~~~\includegraphics[scale=0.33]{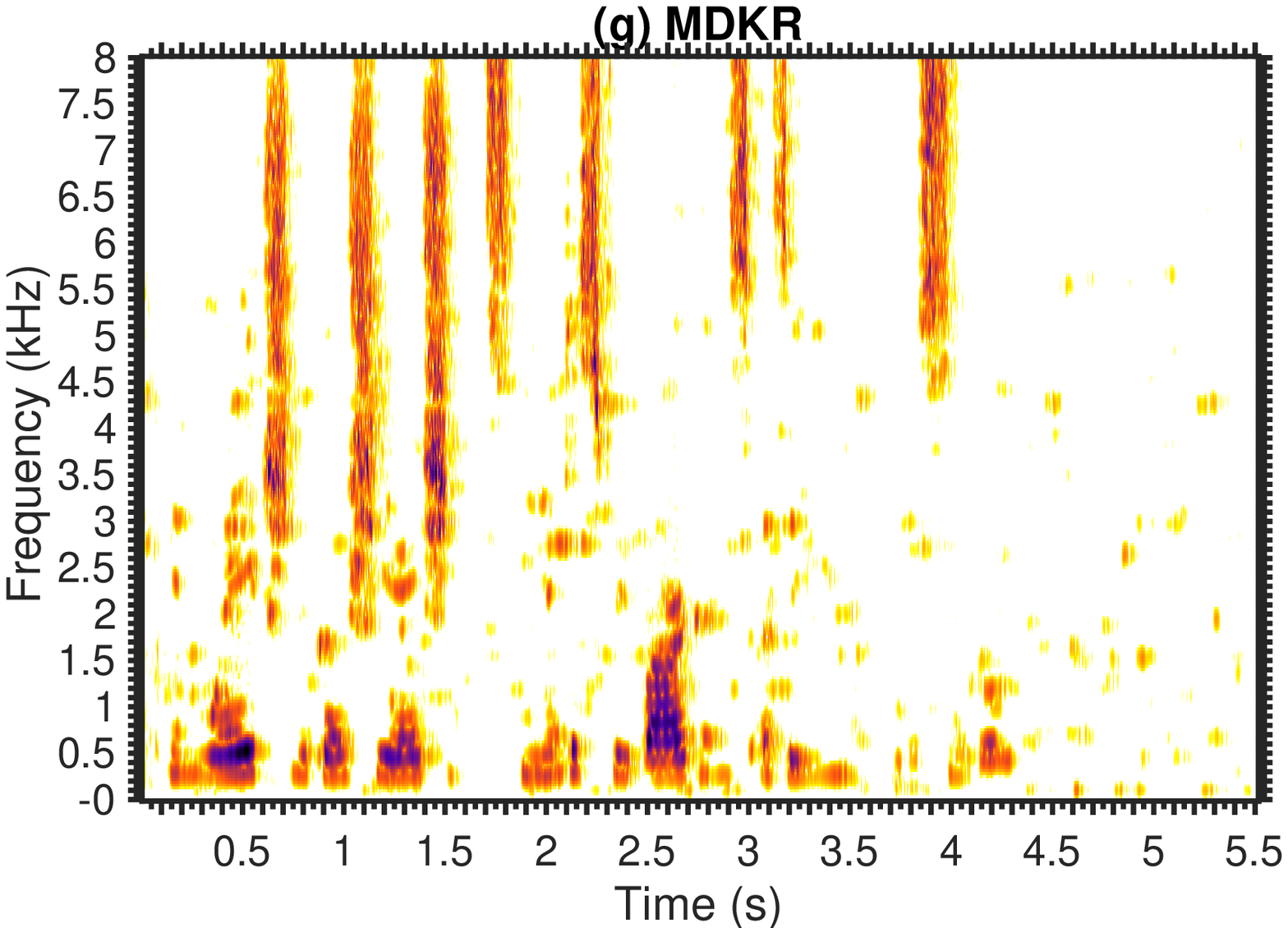}
\par\end{centering}
\caption{Spectrograms of speech enhanced by different enhancers. The noisy
speech was corrupted by F16 noise at 0\,dB SNR.\label{fig:Spectrograms}}
\end{figure*}

In order to assess the performance of the enhancers for speech intelligibility,
the STOI measure \cite{Taal2011} was used. Figures \ref{fig:STOI-Kring-car}
and \ref{fig:STOI-Kring-street} give the average STOI of the noisy
speech and the average STOI performance improvement over noisy speech
at each SNR. It can be seen that for F16 noise, the DNN enhancer performs
better than the other enhancers for SNRs in the range $[-10,10]$\,dB.
At 0\,dB SNR, the DNN enhancer gives an improvement of around 0.015
over the MDKR enhancer; this corresponds to an SNR gain of 0.5 dB.
The MDKR enhancer gives a similar performance to the MDKM and MDKFC
enhancers at high SNRs and it gives an improvement of about 0.01 over
the logMMSE enhancer. For street noise, the DNN enhancer outperforms
other enhancers at SNRs $<$ 10\,dB and at -10\,dB SNR, it gives
an improvement of about 0.035 over the MDKR enhancer which corresponds
to an SNR gain of 2 dB. For SNRs $<$ 5\,dB, the MDKR enhancer outperforms
the MDKM, MDKFC and logMMSE enhancers and at -10\,dB SNR, it gives
an improvement of about 0.018 over the MDKM and about 0.026 over the
MDKFC and logMMSE enhancers. 

In addition to metrics for speech quality and intelligibility, we
have compared the performance of the enhancers on a ASR system trained
on the clean speech signals from the TIMIT dataset. The TMIT core
test set was corrupted by F16 and street noise at 0, 5, 10, 15\,dB
SNRs. A speaker adapted DNN-hidden Markov model (HMM) hybrid system
was trained using the Kaldi toolkit \cite{povey2011kaldi}. The input
features were 40-dimensional feature-space maximum likelihood linear
regression (fMLLR) transformed Mel-frequency cepstral coefficients
(MFCCs). The input context window spanned from 5 frames into the past
to 5 frames into the future. The DNN had 6 hidden layers and around
2000 triphone states were used as the training targets. Initialisation
was performed using restricted Boltzmann machine (RBM) pre-training.
The pre-trained model was then fine-tuned using the frame-level cross-entropy
criterion. Sequence discriminative training using the state-level
minimum Bayes risk (sMBR) criterion \cite{vesely2013sequence} was
then applied. Figures \ref{fig:ASR_f16} and \ref{fig:ASR_street}
give the phone error rate (PER) improvement over noisy speech at each
SNR. It shows that for F16 noise, the MDKR enhancer outperforms other
enhancers at 0, 5 and 10\,dB SNRs. At 0\,dB SNR, the MDKR gives
an improvement of 1\% over the MDKM algorithm and 3\% over the DNN
enhancer. At 15\,dB SNR, the MDKR enhancer performs similarly to
the DNN enhancer and it outperforms the MDKM enhancer by 1\% and the
logMMSE enhancer by 1.7\%. For street noise, the DNN enhancer performs
slightly better than the MDKR enhancer at 0 and 5\,dB SNRs and it
gives an improvement of 2\% over the MDKR and MDKM enhancer. However,
at 10 and 15\,dB, the MDKR enhancer gives similar as the DNN enhancer
and they outperform other enhancers by 0.5\% at 15\,dB SNR. 

\begin{figure*}
\begin{centering}
\includegraphics[scale=0.33]{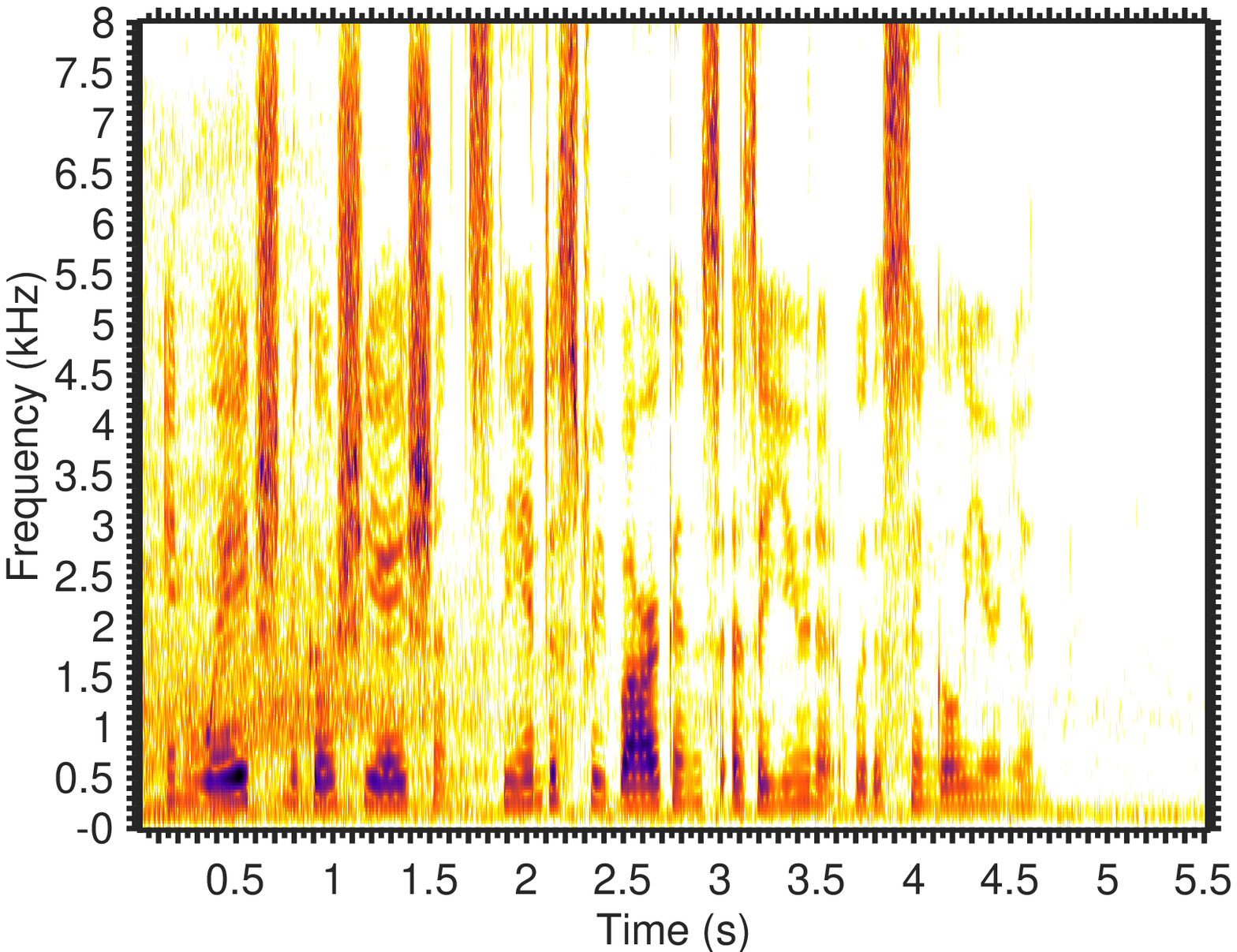}\includegraphics[scale=0.33]{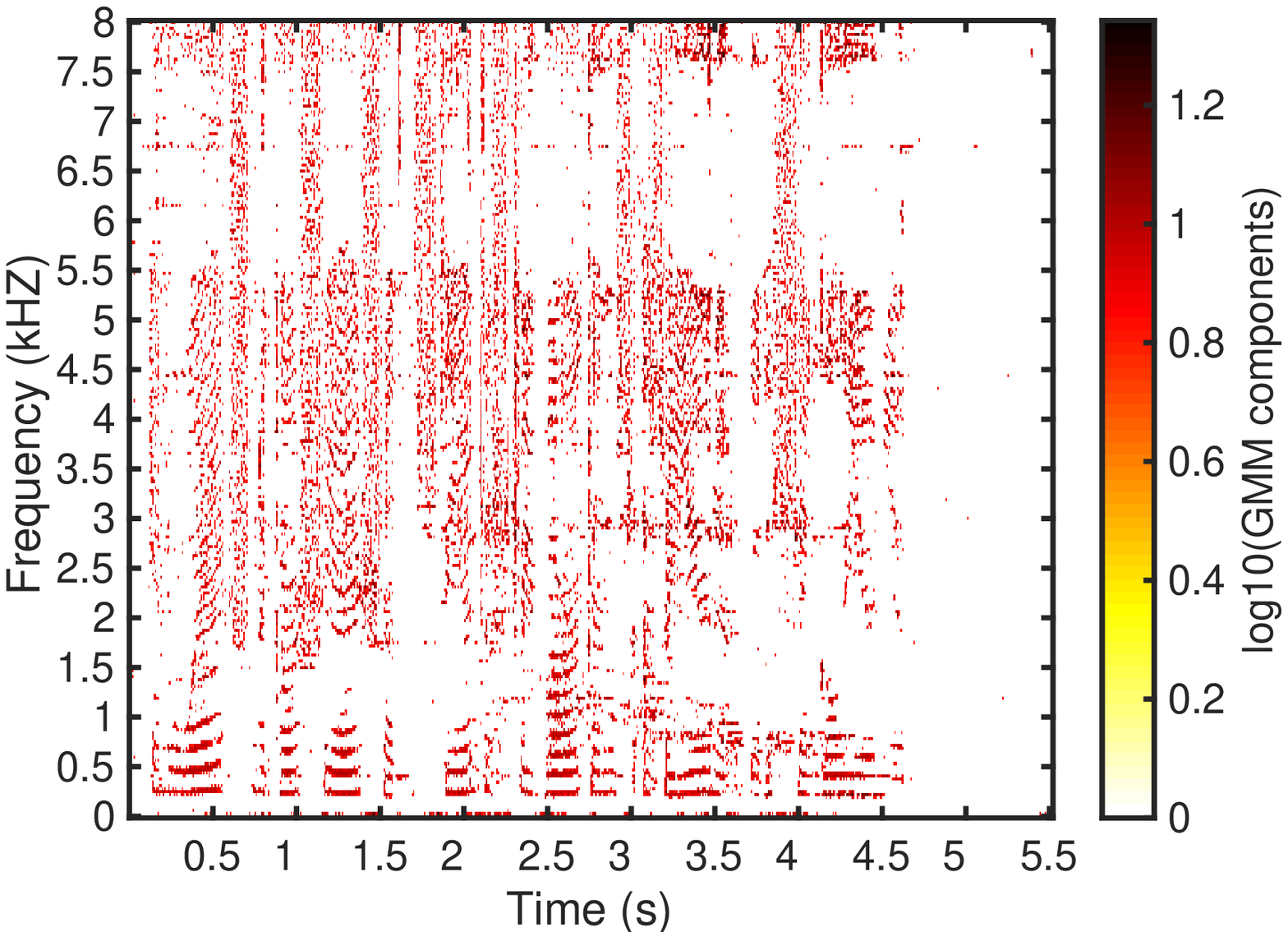}\includegraphics[scale=0.33]{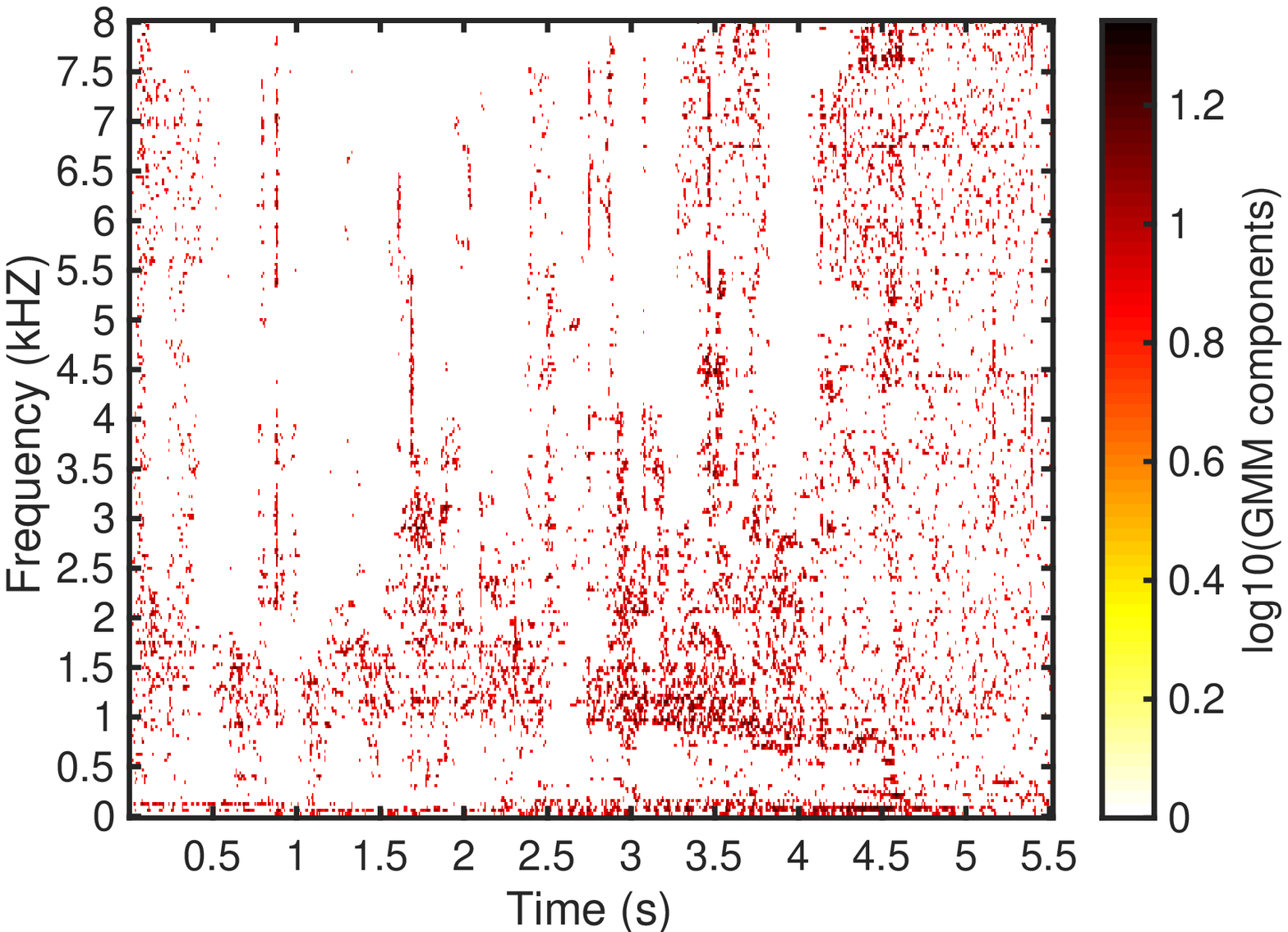}
\par\end{centering}
\caption{Left: Spectrogram of noisy speech at 10\,dB, where the speech is
corrupted by street noise. Middle: number of speech GMM components
for each time-frequency cell. Right: number of noise GMM components
for each time-frequency cell. The numbers of the GMM components have
been transformed into $\text{log}10$ domain for better visualisation.
\label{fig:numgauss_SNR10}}
\end{figure*}

\begin{figure}
\begin{centering}
\includegraphics[scale=0.4]{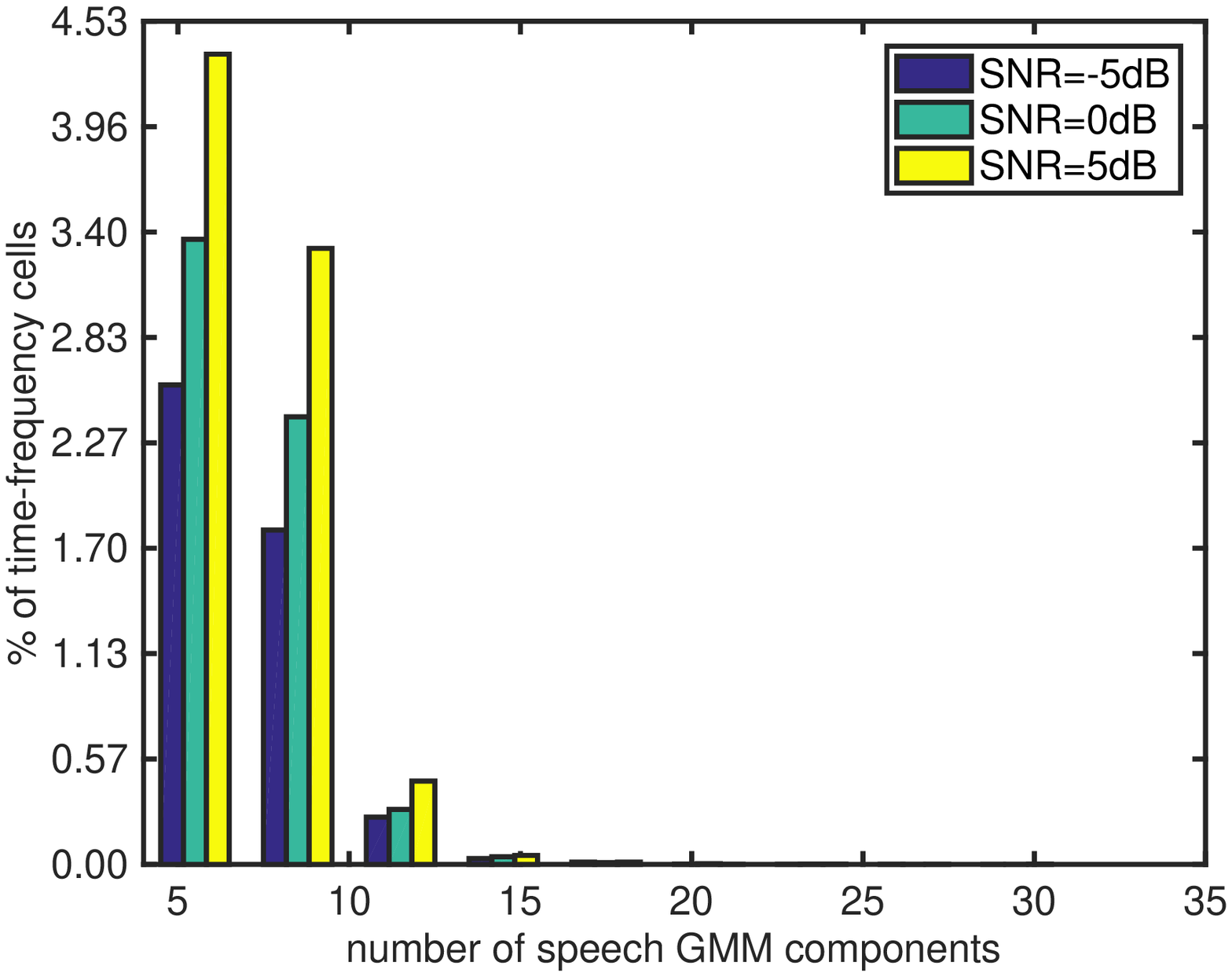}
\par\end{centering}
\begin{centering}
\includegraphics[scale=0.4]{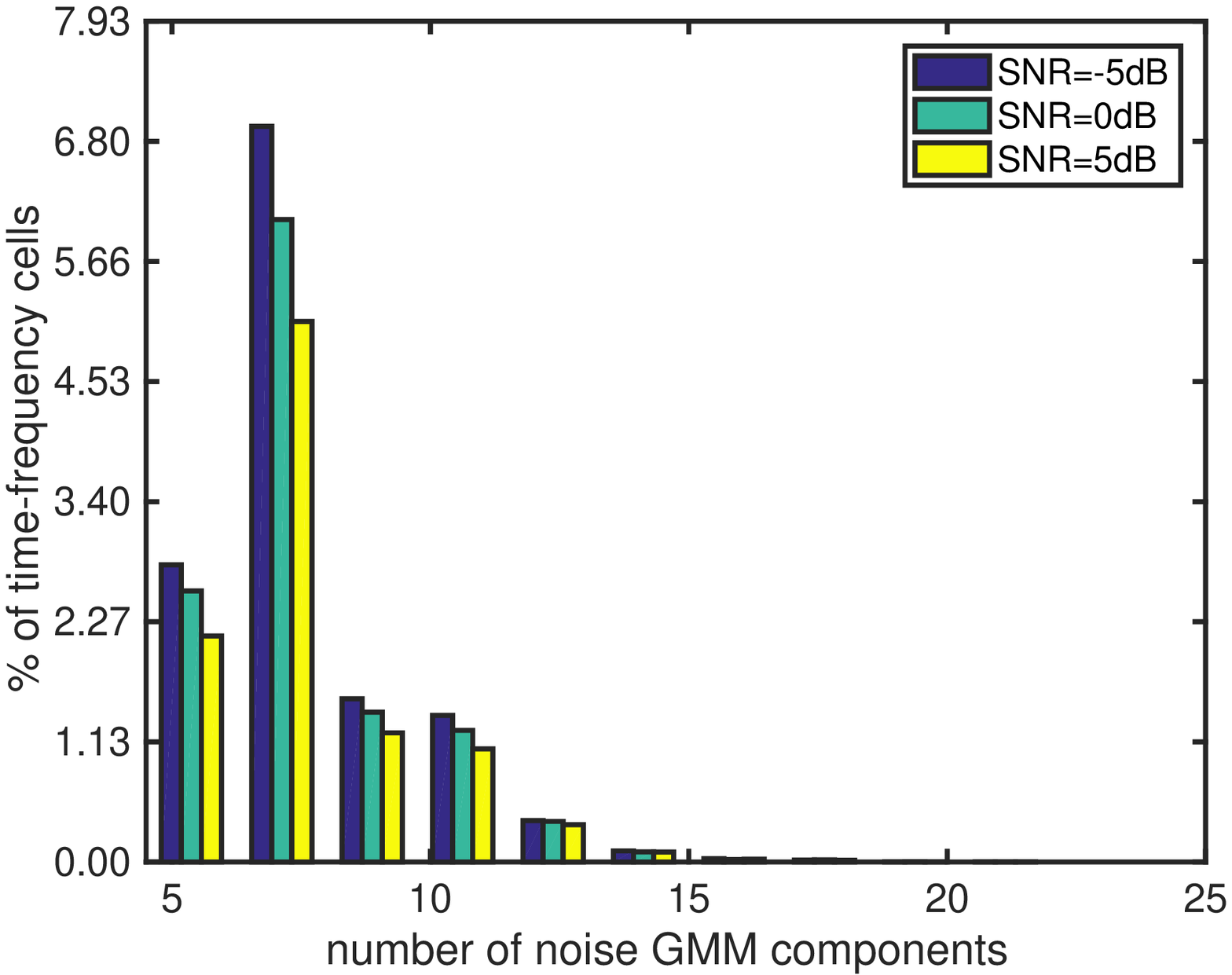}
\par\end{centering}
\caption{Distribution of number of Gaussians components of speech (top) and
noise (bottom) when speech is corrupted by street noise at $-5$,
$0$ and $5$\,dB SNRs.\label{fig:Histogram-Number-of-Gaussians}}
\end{figure}

The spectrograms of speech that has been enhanced by different enhancers
are shown in Fig.~\ref{fig:Spectrograms}. It can be seen that the
MDKR enhancer is better at suppressing noise than other enhancers,
especially in the regions where speech is absent. On the other hand,
the residual noise level of the DNN enhanced speech is higher than
the modulation-domain Kalman filter based enhancers. Compared to the
MDKM and MDKFC enhancers, the MDKR enhancer results in fewer musical
noise artefacts. 

It is interesting to investigate the relationship, for each time-frequency
cell, between the number of Gaussian components chosen by the proposed
Gaussring model and the SNR. In Fig.~\ref{fig:numgauss_SNR10}, the
number of Gaussian components for speech and noise are shown when
the same utterance from Fig.~\ref{fig:Spectrograms}(a) is corrupted
by street noise at 10 dB SNR. For better visualisation, the numbers
of the Gaussian components have been transformed into $\text{log10}$
domain. We can see that for time-frequency cells where the speech
power is high, the predicted speech amplitudes have a high confidence
and thereby the ratio of the prior mean and standard deviation $\frac{\mu_{n|n-1}}{\sigma_{n|n-1}}$
is large. Thus, the speech Gaussring model has a large number of Gaussian
components. Conversely, for time-frequency cells where the noise power
is high, the noise Gaussring model has a large number of Gaussian
components. In Fig.~\ref{fig:Histogram-Number-of-Gaussians}, the
histograms show the distributions of the number of Gaussian components
of speech and noise respectively for speech that is corrupted by street
noise at $-5$, $0$ and $5$\,dB SNRs. When plotting the histograms,
for clarity the histogram plots omit the bars corresponding to $G=1$
(i.e. a single GMM component); these correspond to cells in which
the ratio $\frac{\mu_{n|n-1}}{\sigma_{n|n-1}}<\frac{1}{\sqrt{\frac{4}{\pi}-1}}$
and the Gaussring model backs off to a Rayleigh distribution. It can
be seen that, as the SNR increases, the number of speech components
in each histogram cell increases while the number of noise components
decreases.

\section{Conclusion\label{sec:Conclusion}}

In this paper, a model-based estimator for the spectral amplitudes
of clean speech based on a modulation-domain Kalman filter has been
proposed. The novelty of this proposed enhancer over our previous
work is that it can incorporate the temporal dynamics of both the
speech and noise spectral amplitudes. To obtain the optimal estimate,
a Gaussring model was proposed in which mixtures of Gaussians were
employed to model the prior distribution of the speech and noise in
the complex Fourier domain, leading to the proposed MDKR enhancer.
Over a wide range of SNRs, the MDKR enhancer resulted in enhanced
speech with higher scores for objective speech quality measures than
competing algorithms. For speech intelligibility, the MDKR enhancer
gave worse but yet comparable performance when compared to the DNN
enhancer. The ASR experiments showed that the MDKR enhancer performed
better than competing algorithms for F16 noise and for street noise,
the MDKR enhancer performed similarly to the DNN enhancer for SNRs
$\geq10$\,dB. 

\noindent \bibliographystyle{unsrt}
\bibliography{SAPbib.bib}
\begin{IEEEbiography}[{\includegraphics[width=1in,height=1.25in]{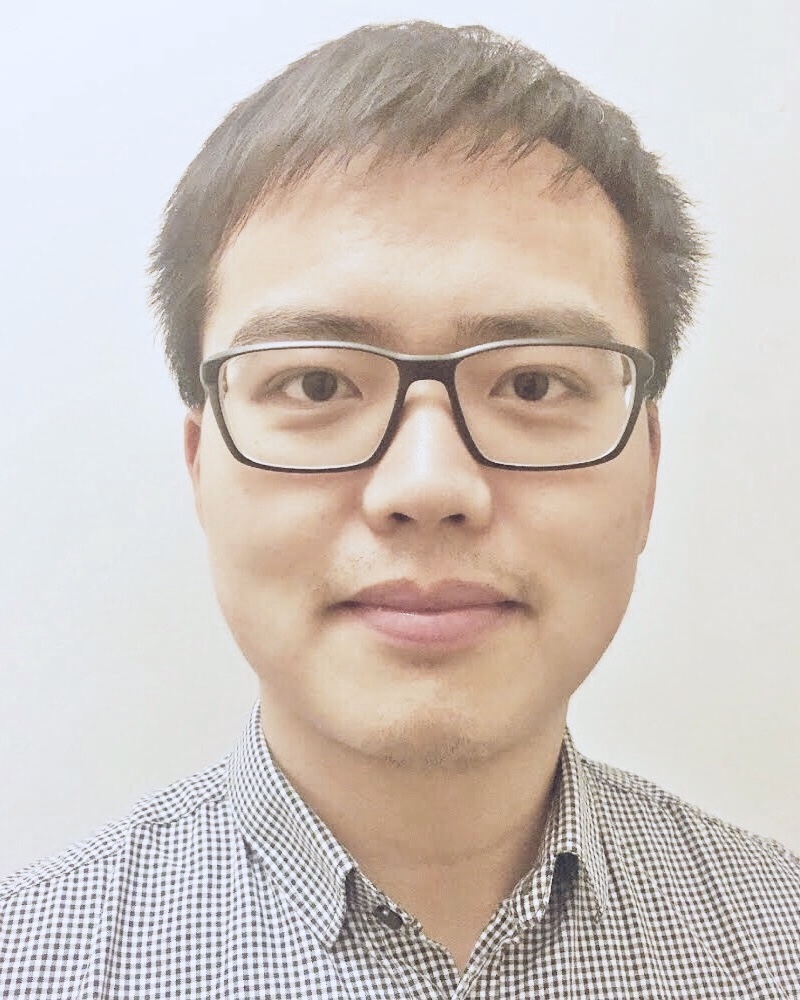}}]{Yu Wang}

(S'12-M'15) received the Bachelor\textquoteright s degree from Huazhong
University of Science and Technology, Wuhan, China, in 2009, the M.Sc.
degree in communications and signal processing and the Ph.D. degree
in signal processing, both from Imperial College, London, U.K. in
2010 and 2015, respectively. Since August 2015 he has been working
as a Research Associate at the Machine Intelligence Laboratory in
the Engineering Department, University of Cambridge. His current research
interests include robust speech recognition, speech and audio signal
processing and automatic spoken language assessment. 

\end{IEEEbiography}

\begin{IEEEbiography}[{\includegraphics[width=1in,height=1.25in]{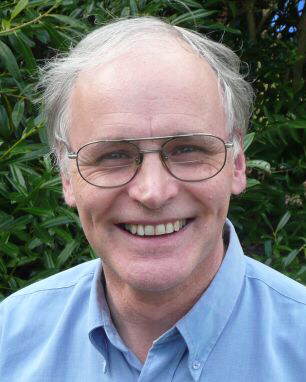}}]{Mike Brookes}

Mike Brookes (M'88) is a Reader (Associate Professor) in Signal Processing
in the Department of Electrical and Electronic Engineering at Imperial
College London. After graduating in Mathematics from Cambridge University
in 1972, he worked at the Massachusetts Institute of Technology and,
briefly, the University of Hawaii before returning to the UK and joining
Imperial College in 1977. Within the area of speech processing, he
has concentrated on the modelling and analysis of speech signals,
the extraction of features for speech and speaker recognition and
on the enhancement of poor quality speech signals. He is the primary
author of the VOICEBOX speech processing toolbox for MATLAB. Between
2007 and 2012 he was the Director of the Home Office sponsored Centre
for Law Enforcement Audio Research (CLEAR) which investigated techniques
for processing heavily corrupted speech signals. He is currently principal
investigator of the E-LOBES project that seeks to develop environment-aware
enhancement algorithms for binaural hearing aids.

\end{IEEEbiography}

\end{document}